\documentclass[sigconf]{acmart}
\AtBeginDocument{  }

\copyrightyear{2026}
\acmYear{2026}
\setcopyright{cc}
\setcctype{by-nc-nd}
\acmConference[CHI '26]{Proceedings of the 2026 CHI Conference on Human Factors in Computing Systems}{April 13--17, 2026}{Barcelona, Spain}
\acmBooktitle{Proceedings of the 2026 CHI Conference on Human Factors in Computing Systems (CHI '26), April 13--17, 2026, Barcelona, Spain}
\acmPrice{}
\acmDOI{10.1145/3772318.3790323}
\acmISBN{979-8-4007-2278-3/2026/04}

\usepackage{multirow}\usepackage{url}
\usepackage{colortbl}
\usepackage{acmart-taps}

\newcommand{\mt}{\mathit{MT}}
\newcommand{\er}{\mathit{ER}}
\newcommand{\sd}{\mathit{SD}}

\newcommand{\id}{\mathit{ID}}
\newcommand{\ide}{\mathit{ID}_\mathrm{e}}
\newcommand{\tp}{\mathit{TP}}

\newcommand{\diff}{\mathit{TP}_\mathrm{diff}}
\newcommand{\cv}{\mathit{TP}_\mathrm{cv}}
\newcommand{\aic}{\mathrm{AIC}}
\newcommand{\bic}{\mathrm{BIC}}
\newcommand{\bias}{\textsf{bias}}
\newcommand{\acc}{\textsf{accurate}}
\newcommand{\neu}{\textsf{neutral}}
\newcommand{\fast}{\textsf{fast}}
\newcommand{\Bias}{\textsf{Bias}}

\newcommand{\xTT}{{\mathit{x}\mathrm{TT}}}
\newcommand{\xCT}{{\mathit{x}\mathrm{CT}}}
\newcommand{\xTTAe}{{\mathit{x}\mathrm{TT}\mathit{A}_\mathrm{e}}}
\newcommand{\xCTAe}{{\mathit{x}\mathrm{CT}\mathit{A}_\mathrm{e}}}
\newcommand{\xyTT}{{\mathit{xy}\mathrm{TT}}}
\newcommand{\xyCT}{{\mathit{xy}\mathrm{CT}}}
\newcommand{\xyTTAe}{{\mathit{xy}\mathrm{TT}\mathit{A}_\mathrm{e}}}
\newcommand{\xyCTAe}{{\mathit{xy}\mathrm{CT}\mathit{A}_\mathrm{e}}}

\begin{document}

\title[Normalizing Speed-accuracy Biases in 2D Pointing Tasks]{Normalizing Speed-accuracy Biases in 2D Pointing Tasks with Better Calculation of Effective Target Widths}

\author{Shota Yamanaka}\email{syamanak@lycorp.co.jp}\orcid{0000-0001-9807-120X}
\affiliation{\institution{LY Corporation}\city{Chiyoda-ku}\state{Tokyo}\country{Japan}}
\author{I. Scott MacKenzie}\email{mack@yorku.ca}\orcid{0000-0003-1731-9651}
\affiliation{\institution{York University}\city{Toronto}\state{Ontario}\country{Canada}}
\renewcommand{\shortauthors}{Yamanaka and MacKenzie}

\begin{abstract}
For evaluations of 2D target selection using Fitts' law, ISO 9241-411 recommends using the effective target width ($W_\text{e}$) calculated using the univariate standard deviation of selection coordinates. Related research proposed using a bivariate standard deviation; however, the proposal was only tested using a single speed-accuracy bias condition, thus the assessment was limited.
We compared the univariate and bivariate techniques in a 2D Fitts' law experiment using three speed-accuracy biases and 346 crowdworkers.  Calculating $W_\text{e}$ using the univariate standard deviation yielded higher model correlations across all bias conditions and produced more stable throughput among the biases.
The findings were also consistent in cases using randomly sampled subsets of the participant data. We recommend that future research should calculate $W_\text{e}$ using the univariate standard deviation for fair performance evaluations.
Also, we found trivial effects when using nominal or effective amplitude and using different perspectives of the task axis.

\end{abstract}



\maketitle

\section{Introduction}
A common research theme in HCI is the evaluation of input devices, interaction techniques, or environmental attributes across computing systems, including smartphones, tablets, virtual reality (VR) systems, and so on.
Gathering and comparing human performance metrics form the bedrock of this research.  To enable fair performance comparisons, evaluation methodologies have standardized many tasks, for example, text entry \cite{Soukoreff03metric,Zhang19text} or path-steering \cite{Accot99,Kasahara23chi}.
To assess human performance in target selection tasks, Fitts' law provides an accepted and well-established protocol. Since 1999 the protocol has been standardized in ISO 9241-411 \cite{iso2012}.
Detailed procedures for experimentation and data analysis are described in other sources (e.g., \cite{mackenzie2018a,Soukoreff04}).

In target selection experiments, participants are typically instructed to perform the task ``as quickly and accurately as possible'' \cite{MacKenzie92,Soukoreff04}.
However, prior work reports that participants may unconsciously bias their behavior toward speed or accuracy \cite{Sharif20,Yamanaka22dis}.
A key issue is that throughput ($\tp$) -- a human performance metric for target selection tasks -- tends to increase when participants prioritize speed \cite{Olafsdottir12test,Yamanaka22chiBias,Batmaz22Improving,Yamanaka24MobileHCI,Batmaz22Effective,Mughrabi23On}.
Thus, even if one group of participants shows higher $\tp$ than another, the difference may simply reflect more reckless pointing rather than truly superior performance.
To compare the performance of devices or participant groups more fairly, an evaluation metric that normalizes the influence of speed and accuracy is needed.

Calculating $\tp$ includes a step that uses the standard deviation ($\upsigma$) of selection endpoints to calculate the effective target width ($W_\mathrm{e}$).
Traditionally, for multi-directional tasks, $\upsigma$ is obtained by projecting endpoints onto the task axis (i.e., the direction of motion for each trial) and computing the univariate standard deviation ($\upsigma_\mathit{x}$) \cite{iso2012,mackenzie2018a,Soukoreff04}.
More recently, Wobbrock et al. proposed that $W_\mathrm{e}$ should be computed using a bivariate standard deviation $(\upsigma_\mathit{xy}$) that also accounts for endpoint variability along the axis perpendicular to the task axis \cite{Wobbrock11dim}.

An observation when using $W_\mathrm{e}$ is that the model fit ($R^2$) remains high even when movement times ($\mt$s) are measured under several speed-accuracy bias conditions \cite{Zhai04speed,Yamanaka24ijhcimerit,Yamanaka22chiBias} and that $\tp$s coalesce across different bias conditions \cite{MacKenzie08,Yamanaka22chiBias,Olafsdottir12test}.
However, for multi-directional pointing tasks, no study compared the model fit and $\tp$ stability when using $\upsigma_\mathit{x}$ and $\upsigma_\mathit{xy}$ across multiple biases.
Because Wobbrock et al. tested only a single bias instruction \cite{Wobbrock11dim}, whether $\upsigma_\mathit{xy}$ yields more stable $\tp$s than $\upsigma_\mathit{x}$ remains unverified.

In view of the above, we conducted an ISO-style evaluation with three speed-accuracy bias instructions to test which computation better normalizes the speed-accuracy bias effect.
As a result, $\upsigma_\mathit{x}$ produced a higher model fit for mixed-bias $\mt$ data and yielded more stable $\tp$s across three bias conditions.
Additionally, even for randomly sampled subsets of participant data, $\upsigma_\mathit{x}$ remained the preferred choice in most cases.
These findings support prior studies that adopted the traditional $W_\mathrm{e}$ computation recommended by ISO 9241-411.
We also compared the effects of using effective amplitude ($A_\mathrm{e}$) and using different task-axis definitions and found that these factors have only minor impact.

\begin{figure*}[t]
\centering
\includegraphics[width=1.0\textwidth]{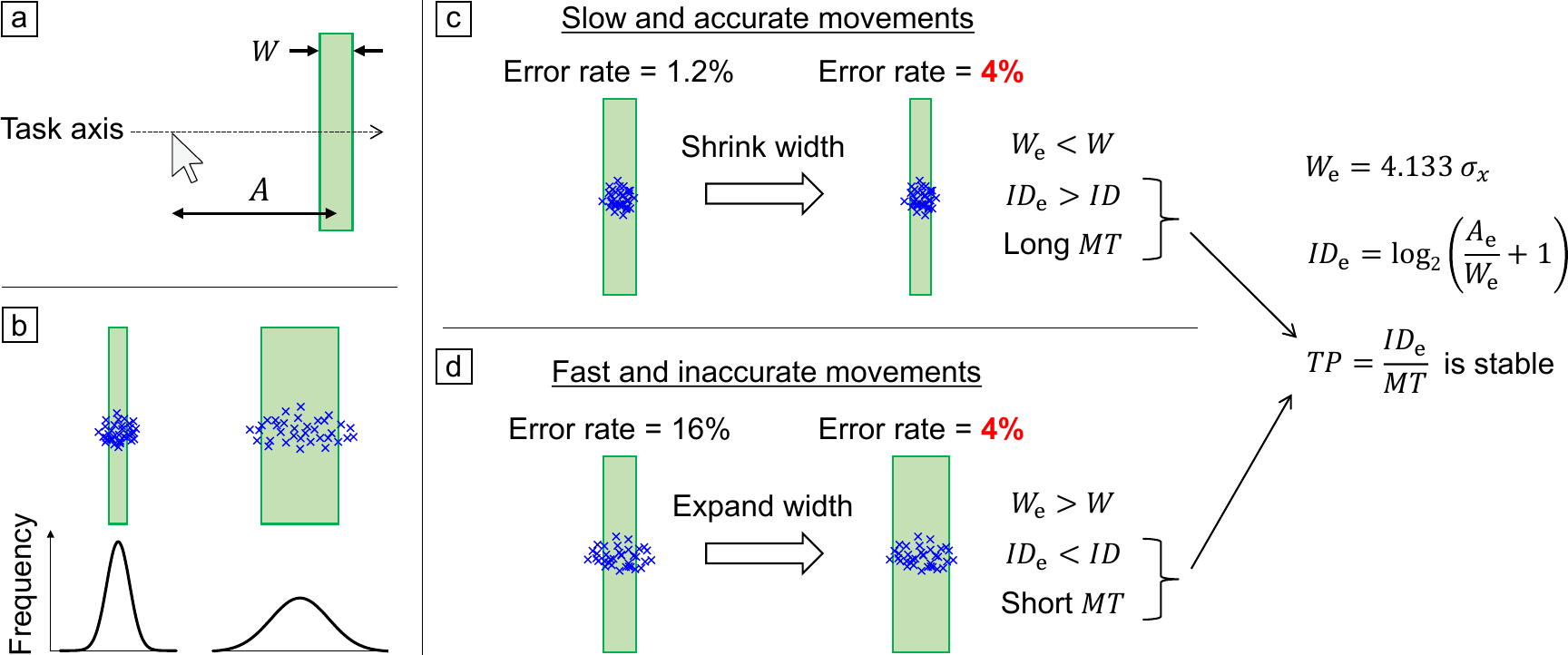}
\caption{The effective-width method for 1D targets. Cross marks (``$\times$'') denote endpoints. (a) In a 1D Fitts' law task where the $x$-axis represents the task axis, a participant aims for a target of width $W$ located at a distance $A$ using a cursor.
(b) When the target is repeatedly selected, the spread of endpoints tends to form a normal distribution, and larger targets generally produce greater endpoint variability.
(c) When participants are biased toward performing in a slow and accurate manner, the error rate is low but the movement time increases.
Conversely, (d) when participants are biased toward fast and inaccurate movements, the error rate increases while the movement time decreases.
By adjusting target width a posteriori, we can compare performance fairly using $\tp$ regardless of accuracy variability.}
\label{fig:introTPstability}
\Description{Diagram of the effective-width method for 1D targets. Panels illustrate how projecting endpoints onto the task axis yields an effective width (We=4.133sigma) that normalizes accuracy so throughput can be compared fairly.}
\end{figure*}

There are three contributions of the work described herein:
(1) By introducing three speed-accuracy bias instructions in 2D pointing tasks, we demonstrate that $\upsigma_\mathit{x}$ normalizes bias effects more than $\upsigma_\mathit{xy}$.
This is novel and advances our understanding of the speed-accuracy tradeoff in HCI.
Our results help correct the tendency of future researchers to follow the use of $\upsigma_\mathit{xy}$.
(2) We perform random sampling to simulate small-sample studies and show that our conclusions holds in most cases.
This addresses a common issue in HCI research where large samples are not available, and highlights the robustness of $\upsigma_\mathit{x}$ for future research.
(3) We provide implications for researchers on the model to choose when conducting Fitts' law experiments comparing devices, interaction techniques, environmental attributes, or participant groups.

\section{Related Work}
\subsection{Fitts' Law and the Speed-Accuracy Tradeoff}

According to Fitts' law, the movement time ($\mathit{MT}$ in seconds) to point to and acquire a target is linearly related to a task's index of difficulty ($\mathit{ID}$ in bits):

\begin{equation}
\label{eqn:defFittsLaw1}
\mt=a+b\cdot{}\id\ 
\end{equation}

\noindent with

\begin{equation}
\label{eqn:defFittsLaw2}
\id = \log_{2}\left(\frac{A}{W}+1\right).
\end{equation}

\noindent See Figure~\ref{fig:introTPstability}a. Above, $A$ is the movement amplitude, $W$ is the target width, and $a$ and $b$ are empirical constants.
$\id$ is a nominal value computed from $A$ and $W$ in screen coordinates, such as pixels or cm.
In typical experiments, participants are instructed to perform ``as quickly and accurately as possible'' \cite{MacKenzie92,Soukoreff04}.
We refer to this speed-accuracy balancing instruction as a $\neu$ bias \cite{Zhai04speed,Yamanaka22chiBias}.

Previous studies used pointing tasks to compare performance across input devices \cite{Card78,Jones2020Leap} and user attributes \cite{Welford69age,Ren11,Ling2024Model}.
However, if participants perform in one condition quickly with a high error rate ($\er$) and another condition slowly with a low $\er$, we cannot judge the former superior solely because of its lower $\mt$.
If the two conditions were performed at the same $\er$, the device with lower $\mt$ would be considered better, but such a coincidence is rare in practice.

To enable fair comparisons, Crossman's correction is recommended \cite[pp.~75--76]{Crossman56} \cite[p.~110]{fitts1964a}.
The method adjusts the target size based on the variability in the selection coordinates over a sequence of trials.
Given this, an effective index of difficulty ($\ide$) is defined as

\begin{equation}
\label{eqn:defWe1}
\ide = \log_{2}\left(\frac{A_\mathrm{e}}{W_\mathrm{e}}+1\right)
\end{equation}

\noindent with

\begin{equation}
\label{eqn:defWe2}
W_\mathrm{e}=\sqrt{2 \uppi e}\upsigma = 4.133\upsigma.
\end{equation}

\noindent Above,
$W_\mathrm{e}$ is the effective target width, $\upsigma$ is the standard deviation of endpoints, and 
$A_\mathrm{e}$ is the effective or mean movement amplitude over a sequence of trials.
The reader is referred to other sources for full details on the theory and practice for Equation~\ref{eqn:defWe2} (\cite[pp.~106--109]{MacKenzie92} \cite[pp.~355--358]{mackenzie2018a} \cite[pp.~147--148]{welford1968a}).

For 1D tasks with horizontal movement, $\upsigma$ is the standard deviation along the $x$-axis.
The adjustment $W_\mathrm{e} = 4.133\upsigma$ assumes that the endpoints are normally distributed about the target. See Figure~\ref{fig:introTPstability}b.
This adjusts $W_\mathrm{e}$ so that 3.88\% ($\approx$4\%) of clicks fall outside the target.
We then obtain $\tp$, which is a metric that fairly compares performance under the same (albeit adjusted) $\er$, regardless of whether the task is performed slowly or quickly. See Figure~\ref{fig:introTPstability}c and Figure~\ref{fig:introTPstability}d, respectively.  $\tp$ (in bits per second or bps) for each participant is calculated as

\begin{equation}
\tp = \frac{1}{k} \sum\nolimits_{i=1}^{k} \left( \frac{\id_{\mathrm{e}_i}}{\mt_{i}} \right), 
\label{eqn:defTP}
\end{equation}

\noindent where $k$ is the number of $A$-$W$ target conditions, and $i$ indexes the $i$-th condition.
The validity of the effective-width method is also supported by Gori et al. who found that both nominal $\id$ and $\ide$ can be derived from cursor velocity profiles \cite{Gori20BC,Gori23pvp}.

Another version of $\tp$ uses the reciprocal of the regression slope in a Fitts' law model ($\frac{1}{b}$ from Equation~\ref{eqn:defFittsLaw1}) \cite{Wobbrock11ART,Zhai04chara,Fitts54}, but problems emerge due to ignoring the intercept \cite{Soukoreff04,Yamanaka24MobileHCI}.
We therefore consistently use the definition in Equation~\ref{eqn:defTP} and include results for the slope-reciprocal $\tp$ in the supplementary materials.

\subsection{Influence of Effective Width 1: Improving Model Fit Across Bias Conditions}

Inserting speed-accuracy biases in pointing tasks dates to Fitts and Radford \cite{Fitts66}.
They used monetary incentives and penalties, while other research used verbal instructions \cite{Zhai04speed,Guiard11explicit}, text-based instructions \cite{Yamanaka24ijhcimerit,Yamanaka22chiBias}, or metronomes \cite{Wobbrock08error,Wobbrock11error}.
These studies reported that emphasizing speed increases $\er$, and conversely emphasizing accuracy reduces $\er$.

One effect of using $W_\mathrm{e}$ to normalize the speed-accuracy bias is that regressing $\mt$ data across bias conditions yields an improved fit compared to using nominal $W$.
Zhai et al. conducted four experiments in which participants performed 1D pointing under different biases \cite{Zhai04speed}.
In Experiment 1, they used three subjective biases: $\neu$, $\acc$ (emphasizing accuracy), and $\fast$ (emphasizing speed).
They regressed $\mt$ across all biases, which they called the ``mixed'' condition.
Using nominal $\id$ yielded $R^2 = .696$, whereas using $\ide$ yielded $R^2 = .825$.
The superiority of $\ide$ in the mixed condition was replicated in their subsequent experiments.

Yamanaka conducted a 1D pointing experiment with 226 crowdworkers and likewise found that $\ide$ provided a higher model fit than nominal $\id$ in the mixed bias condition \cite{Yamanaka24ijhcimerit}.
Randomly sampled smaller sets of participant data (e.g., 10 or 20) confirmed that $\ide$ robustly outperformed nominal $\id$.

Strictly speaking, Fitts' law models $\mt$ under a single condition, and thus fitting to mixed data across multiple instructions is potentially out of scope.
Nevertheless, Zhai et al. \cite{Zhai04speed} and Yamanaka et al. \cite{Yamanaka24ijhcimerit,Yamanaka22chiBias,Yamanaka24MobileHCI} analyzed why $\ide$ still yields a higher fit in the mixed condition:  
(1) As instructions emphasize speed, $W_\mathrm{e}$ increases and $\ide$ appropriately decreases, and vice versa.
(2) Consequently, the per-bias intercept and slope in the $\mt$ regressions become closer, so $\mt$ data points from multiple biases are nearly aligned along a single regression line.
Following this approach, we judge which model (nominal $\id$ vs.\ $\ide$) better normalizes bias via a higher $R^2$ in the mixed condition.

\subsection{Influence of Effective Width 2: Increasing Throughput Stability}

MacKenzie and Isokoski conducted a 1D pointing task with three speed-accuracy biases and reported that using $W_\mathrm{e}$ coalesced the three $\tp$s, ranging from 5.67~bps to 5.73~bps (i.e., a difference below 1\%).
Previous studies with rectangular targets \cite{Yamanaka22chiBias}, finger touch \cite{Yamanaka24MobileHCI}, and goal crossing \cite{Kasahara23chi} used three biases and also reported that Fitts' law or its variants yield more stable $\tp$s with $W_\mathrm{e}$ than with nominal $W$.

In contrast, Olafsdottir et al. tested two additional conditions called ``max speed'' and ``max accuracy''. 
They reported that $\tp$ ranged from 6~bps to 10~bps (a 42\% spread) and argued that $W_\mathrm{e}$ does not normalize the influence of bias \cite{Olafsdottir12test}.
However, they used extreme instructions, for example, under max accuracy, participants were asked to click a 1-pixel-wide target on every trial.
For typical HCI studies comparing devices or user groups, tasks of moderate difficulty and instructions are used, and thus their results do not necessarily refute the utility of $W_\mathrm{e}$.

\subsection{Univariate vs.\ Bivariate Standard Deviation for Effective Width}

\begin{figure*}[t]
\centering
\includegraphics[width=1.0\textwidth]{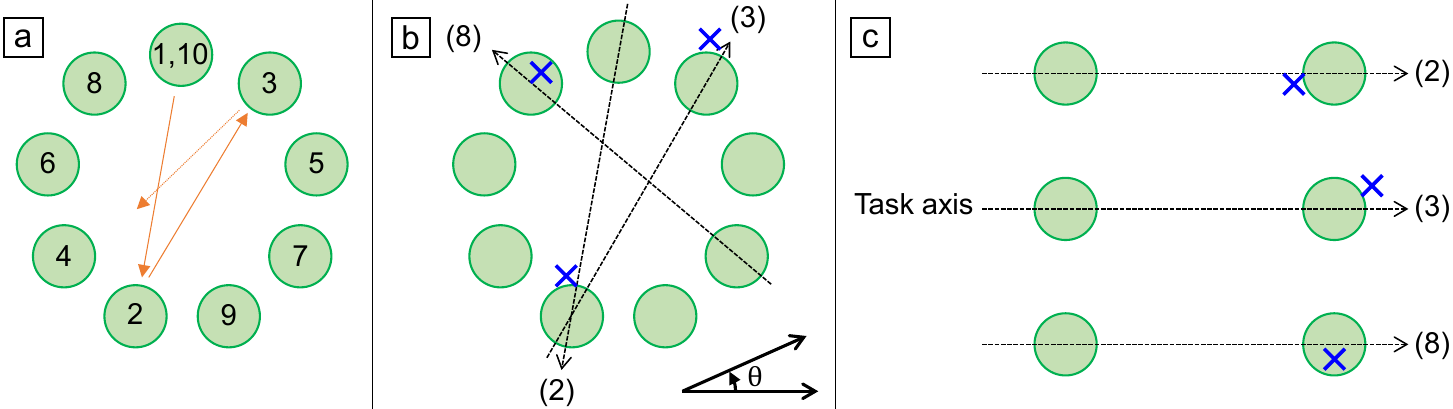}
\caption{(a) ISO 9241-411 multi-directional pointing with circular targets. When there are nine targets, select them in order from 1 to 10. (b--c) Rotating each trial so that the task axis points to the right when computing endpoint distributions.}
\label{fig:rwWobbrockSDxy}
\Description{ISO multi-directional pointing setup with circular targets on a layout circle and the rotation used to define the task axis. Panels show how trials are rotated so the previous-to-current target direction becomes +x for endpoint analysis.}
\end{figure*}

In realistic user interfaces, we often perform multi-directional movements to acquire 2D targets.
To measure performance in such situations, ISO 9241-411 \cite{iso2012} and Soukoreff and MacKenzie \cite{Soukoreff04} recommend selecting a sequence of targets arranged around a layout circle (Figure~\ref{fig:rwWobbrockSDxy}a).
For the standard deviation used in $W_\mathrm{e}=4.133\upsigma$ (Equation~\ref{eqn:defWe2}), those sources \cite{iso2012,Soukoreff04} recommend univariate $\upsigma_\mathit{x}$.
This computation considers only the endpoint spread along the task axis. 
See Figure~\ref{fig:rwWobbrockSDxy}b--c.
Specifically, for $n$ trials under a target condition, let $x_j$ ($j=1,2,\ldots,n$) be the signed distance of each endpoint projected onto the task axis from the target center.
Given this,

\begin{equation}
\mathrm{Univariate\ standard\ deviation\ } \upsigma_\mathit{x} = \sqrt{\frac{1}{n-1} \sum\nolimits_{j=1}^{n} (x_j - \overline{x})^2}.
\end{equation}

\begin{figure}[t]
\centering
\includegraphics[width=1.0\columnwidth]{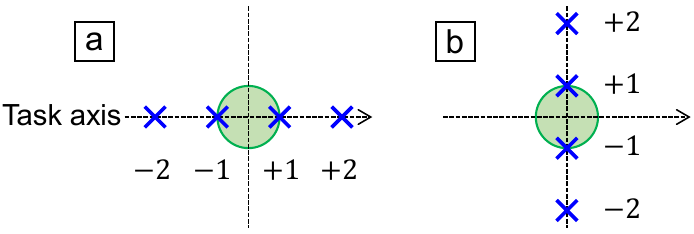}
\caption{Dispersion in the plane is identical, but we have (a) $\upsigma_\mathit{x}=1.83$ and (b) $\upsigma_\mathit{x}=0$ (arbitrary units such as cm or pixels).}
\label{fig:rwWobbrockZeroSDx}
\Description{Counterexample motivating bivariate sigma. Two point sets have identical planar spread, but one has zero univariate sigma_x along the task axis while the other does not; this illustrates a limitation of sigma_x in edge cases.}
\end{figure}

Wobbrock et al. argued that $\upsigma_\mathit{x}$ may not appropriately capture user behavior \cite{Wobbrock11ART}.
For example, consider four clicks as in Figure~\ref{fig:rwWobbrockZeroSDx}, where endpoints are equally dispersed either along the task axis or orthogonal to it.
For the univariate definition, the latter yields $\upsigma_\mathit{x}=0$.
The bivariate standard deviation addresses this issue:  
\aptLtoX{\begin{equation}
\mathrm{Bivariate\ standard\ deviation\ } \upsigma_\mathit{xy} = \sqrt{ \frac{1}{n-1} \sum_{j=1}^{n} \left\{ (x_j - \overline{x})^2 + (y_j - \overline{y})^2 \right\} }.
\end{equation}}{\begin{equation}
\scalebox{0.92}{$
\mathrm{Bivariate\ standard\ deviation\ } \upsigma_\mathit{xy} = \sqrt{ \frac{1}{n-1} \sum_{j=1}^{n} \left\{ (x_j - \overline{x})^2 + (y_j - \overline{y})^2 \right\} }.
$}
\end{equation}}

All trials are rotated so that the task axis ($x$-axis) turns towards the right (Figure~\ref{fig:rwWobbrockSDxy}b--c), and the $y$-axis is orthogonal to it.
$y_j$ is defined analogously to $x_j$.
With this definition, both cases in Figure~\ref{fig:rwWobbrockZeroSDx} yield $\upsigma_\mathit{xy}=1.83$.

\subsection{Studies Citing Wobbrock et al.'s CHI 2011 Paper}
\label{sec:rwCiteWobbrock2011}

We performed a forward citation search (i.e., snowballing with a depth of 1) using the ACM Digital Library and Google Scholar for papers citing Wobbrock et al.'s work \cite{Wobbrock11dim} and found 112 unique items (retrieved August 1, 2025).
Among these, 28 conducted tasks using circular targets or 2D layouts in 3D space.
There were studies that computed $\tp$ using $\upsigma_\mathit{x}$ \cite{Batmaz22Effective,Abe2025Understanding,Kohli2013Redirected}, some that used $\upsigma_\mathit{xy}$ to evaluate Fitts' law variants for touchscreens \cite{Yamanaka18mobilehci,Yamanaka18iss,Yamanaka21imwut}, and others that assessed $\er$ prediction models \cite{Wobbrock11error,Yamanaka20issFFF,Bi16}.

Many papers reporting model fit or $\tp$ used only $\upsigma_\mathit{xy}$ for 2D targets, particularly in comparative studies across conditions.
Table~\ref{tab:studies_sigma_xy} summarizes the survey results.
Unless bias comparison was the goal, only a single bias of $\neu$ was used.
A study not included in the table is Kuberski and Gafos, who used trivariate $\upsigma_\mathit{xyz}$ with intraoral tongue-based operation to compute $W_\mathrm{e}$ and model fit \cite{Kuberski2021Fitts}.

\begin{table*}[t]
\centering
\caption{Studies reporting model fit or $\tp$ using only $\upsigma_\mathit{xy}$ for 2D targets.}
\label{tab:studies_sigma_xy}
\begin{tabular}{llcc}
\hline
\textbf{Comparison category} & \textbf{Comparison targets} & \textbf{Sample size} & \textbf{Reference} \\
\hline
\multirow{5}{*}{Input device} 
& Mouse and Leap Motion & 48 & \cite{Jones2020Leap} \\
& Mouse sensor positions & 14 & \cite{Kim2020Optimal} \\
& Mouse and motion capture-based touchpad & 6 & \cite{Berard2024Congruent} \\
& Cursor's spatiotemporal jitter levels & 13 & \cite{Ham2021FasterMouse} \\
& Mouse, stylus, freehand input, and motion-based controller & 20 & \cite{Habibi2021Handedness} \\
\hline
\multirow{5}{*}{Interaction technique} 
& Interaction areas, e.g., on smartphone screen vs. user defined space & 12 & \cite{Winkler2012Investigating} \\
& Sensors (e.g., Kinect vs. camera) and confirmations (e.g., blink vs. dwell) & N/A & \cite{Guness2013Novel} \\
& Calibration methods for mid-air gestures & 15 & \cite{Guinness2015Models} \\
& Mid-air gestures (head, freehand, hand holding cup) & 12 & \cite{Clarke2017MatchPoint} \\
& On-screen operations (e.g., tap vs. drag) & 20 & \cite{Ng2016Effects} \\
\hline
\multirow{2}{*}{Environment} 
& Laboratory vs. field study & 12 & \cite{Evans2012Taming} \\
& User engagement strategies (e.g., background music, animation) & 21 & \cite{Yu2020Engaging} \\
\hline
Feedback & Pitch variations in sound feedback & 52 & \cite{Lock2020Experimental} \\
\hline
Instruction & Speed-accuracy biases & 16 & \cite{Yamanaka24MobileHCI} \\
\hline
User attribute & With vs. without Parkinson's disease & 40 & \cite{Ling2024Model} \\
\hline
\end{tabular}
\Description{Survey of studies that reported model fit or throughput using only bivariate sigma for 2D targets. Rows summarize comparison focus, sample size, and references across devices, techniques, environments, feedback, instructions, and user attributes.}
\end{table*}

Examples comparing model fit or $\tp$ using both $\upsigma_\mathit{x}$ and $\upsigma_\mathit{xy}$ were rare after Wobbrock et al. \cite{Wobbrock11dim}: We found only four \cite{YU19siggraph,Parisay2021EyeTAP,Parisay2021IDEA,Yamanaka22chiBias}.
Yu et al. conducted ISO-style tasks with spherical targets in VR, using raycasting: head movement in Experiment 1 and handheld controller in Experiment 2 \cite{YU19siggraph}.
Although only $\neu$ bias was used, they reported model fits with nominal $\id$, $\upsigma_\mathit{x}$-based $\ide$, and $\upsigma_\mathit{xy}$-based $\ide$.
In Experiment 1, the respective $R^2$s were .96, .87, and .89, and in Experiment 2 they were .92, .98, and .98.
Thus, $\upsigma_\mathit{x}$ did not outperform $\upsigma_\mathit{xy}$.
However, Yu et al. excluded the longest $A$ condition from analysis in Experiment~2 and noted the importance of considering factors unique to VR such as hand tremor and depth perception.

Parisay et al. used Wobbrock et al.'s FittsStudy tool \cite{Wobbrock11dim} to compare mouse input, gaze with dwell, and speech recognition \cite{Parisay2021EyeTAP,Parisay2021IDEA}.
They reported $\tp$s using both $\upsigma_\mathit{x}$ and $\upsigma_\mathit{xy}$ and consistently obtained higher $\tp$ with $\upsigma_\mathit{x}$.
Comparative conclusions were consistent with both calculations (e.g., the mouse outperformed speech recognition).

Yamanaka et al. used two rectangular targets (width $W$, height $H$) placed horizontally and participants alternately clicked them \cite{Yamanaka22chiBias}.
In a laboratory study with three biases, they reported model fit in the mixed condition.
Nominal $\id$ using only $W$ yielded $R^2=.6151$, whereas $\upsigma_\mathit{x}$ yielded $R^2=.8697$ and $\upsigma_\mathit{xy}$ yielded $R^2=.7507$.
Among the three, $\upsigma_\mathit{x}$ also produced the best $\tp$ stability across biases.
Their follow-up crowdsourcing study reached the same conclusion.

\section{Limitations of Previous Studies and Our Research Questions}

\subsection{Unexamined Aspect 1: The Bias-normalizing Capability of Univariate vs.\ Bivariate \texorpdfstring{$\upsigma$}{sigma}}

Before Wobbrock et al. \cite{Wobbrock11dim}, the recommendations by ISO 9241-411 \cite{iso2012} and Soukoreff and MacKenzie \cite{Soukoreff04} were to use $\upsigma_\mathit{x}$, but we found no empirical research comparing $\upsigma_\mathit{x}$ with $\upsigma_\mathit{xy}$.
To date, there are no investigations on which calculation better realizes the capability of effective width for ISO-style tasks.
Specifically, although Wobbrock et al. \cite{Wobbrock11ART}, Parisay et al. \cite{Parisay2021EyeTAP,Parisay2021IDEA}, and Yu et al. \cite{YU19siggraph} tested both methods, they used only $\neu$ instructions.
Even before Wobbrock et al., device comparisons using $\tp$ computed with $\upsigma_\mathit{xy}$ \cite{Douglas99} and model fits using bivariate joint probability distributions \cite{Murata99} were reported, but again only under a single bias.

To claim that $\upsigma_\mathit{xy}$ better normalizes speed-accuracy bias requires testing model fits in mixed bias conditions \cite{Zhai04speed} and examining the $\tp$ stability across biases \cite{MacKenzie08}.
In particular, Wobbrock et al. recommended $\upsigma_\mathit{xy}$ primarily based on model-fit results under the $\neu$ bias (see Section \ref{sec:why} for further details).
But without multiple biases, the utility of $W_\mathrm{e}$ is not appropriately assessed, casting doubt on that recommendation.

As an exception, Yamanaka et al. compared $\upsigma_\mathit{x}$ and $\upsigma_\mathit{xy}$ under three biases \cite{Yamanaka22chiBias}, but they used rectangular targets and a traditional left-right alternating movements.
Realistic user interfaces require multi-directional movements, and recent work predominantly uses ISO-style targets around a layout circle.
Identifying a method that better normalizes bias in such tasks would thus contribute to HCI.

\subsection{Unexamined Aspect 2: Nominal vs.\ Effective Amplitude}
\label{sec:rq2}

We identified two additional unexamined aspects regarding the effective-width method.
First, the choice between nominal $A$ and $A_\mathrm{e}$ is inconsistent across studies.
Even among papers citing Wobbrock et al. \cite{Wobbrock11dim}, some used $A_\mathrm{e}$ \cite{Guinness2015Models,Winkler2012Investigating,Kohli2013Redirected,Batmaz22Effective,YU19siggraph,Clarke2017MatchPoint}, while others and ISO 9241-411 used $A$ \cite{Guness2013Novel,Ng2016Effects,Yamanaka24MobileHCI,Lock2020Experimental,Ling2024Model,iso2012}.
Zhai et al. argued that ``Since the difference of $D$ vs. $D_\mathrm{e}$ [note: $A$ and $A_\mathrm{e}$ in our paper] is comparatively smaller than that of $W$ vs. $W_\mathrm{e}$, this study focuses on the latter,'' but no evidence was provided \cite{Zhai04speed}.
We therefore compare variants using $A$ and $A_\mathrm{e}$ for both $\upsigma_\mathit{x}$ and $\upsigma_\mathit{xy}$.

\subsection{Unexamined Aspect 3: Definitions of the Task Axis}

\begin{figure}[t]
\centering
\includegraphics[width=1.0\columnwidth]{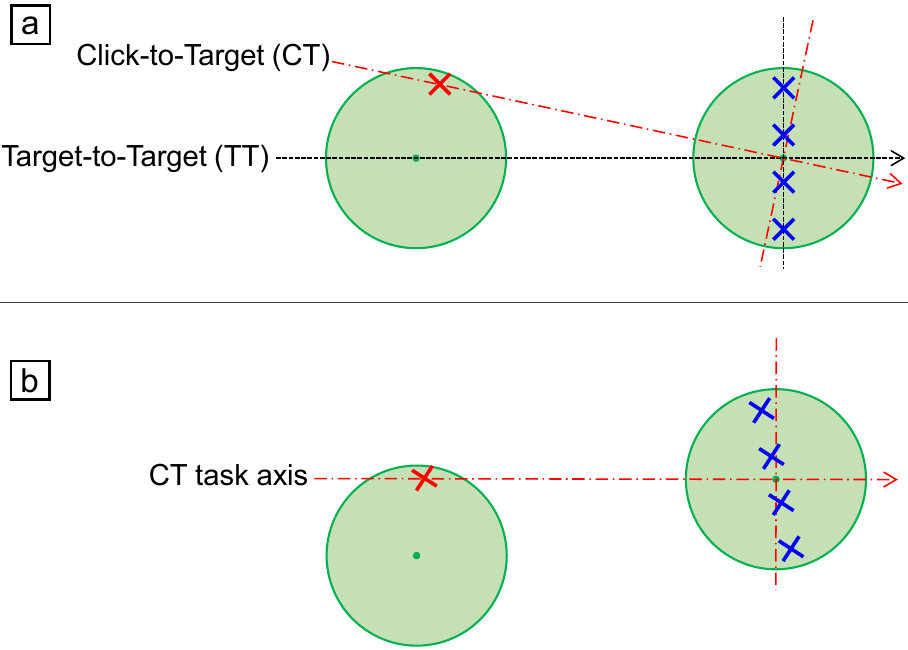}
\caption{Two definitions of the task axis. Depending on whether the start point is the previous target center or the previous successful click position (red cross), TT and CT yield different $\upsigma_\mathit{x}$ values.}
\label{fig:rqTaskAxis}
\Description{Two task-axis definitions for multi-directional pointing. (a) Target-to-target (TT) uses the centers of consecutive targets; (b) Click-to-target (CT) starts from the previous successful click, which can yield a different sigma_x.}
\end{figure}

The second aspect is how to define the task axis.
Representative work by Soukoreff and MacKenzie \cite{Soukoreff04} and Wobbrock et al. \cite{Wobbrock11dim} defined the task axis as the direction from the center of the previous target to the center of the current target (Figure~\ref{fig:rqTaskAxis}a).
We call this the target-to-target (TT) task axis.
We found that the rationale for the TT task axis has not been articulated, but many subsequent studies have followed this tradition.

Soukoreff and MacKenzie described the task axis as ``in the direction of motion'' \cite{Soukoreff04}, and ISO 9241-411 provides a similar explanation \cite{iso2012}.
This admits an alternative interpretation: Instead of the previous target's center, the task axis begins at the cursor position when a trial begins; that is, at the click position on the previous target.
We call this the click-to-target (CT) task axis.  See Figure~\ref{fig:rqTaskAxis}.
For $\upsigma_\mathit{x}$, TT yields $\upsigma_\mathit{x}=0$ in Figure~\ref{fig:rqTaskAxis}a, whereas CT yields $\upsigma_\mathit{x}>0$ in Figure~\ref{fig:rqTaskAxis}b.

The difference between TT and CT may be small.
Because a miss must be re-aimed until success in our experiment, the CT start point lies within the target and would coincide with the target center on average.
If so, the bias-normalizing capability may be similar for TT and CT.
However, we lack evidence that adopting either TT or CT is equally acceptable.
It is valuable for methodological standardization to determine whether researchers should avoid one or whether both are fine.

In summary, we pursued three research questions.  Which conditions should be used to normalize the influence of speed-accuracy bias for fair comparisons:  (RQ1) univariate $\upsigma$ vs. bivariate $\upsigma$, (RQ2) $A$ vs.\ $A_\mathrm{e}$, and (RQ3) TT vs.\ CT task axes?  

\section{Method}

\begin{figure*}[t]
\centering
\includegraphics[width=1.0\textwidth]{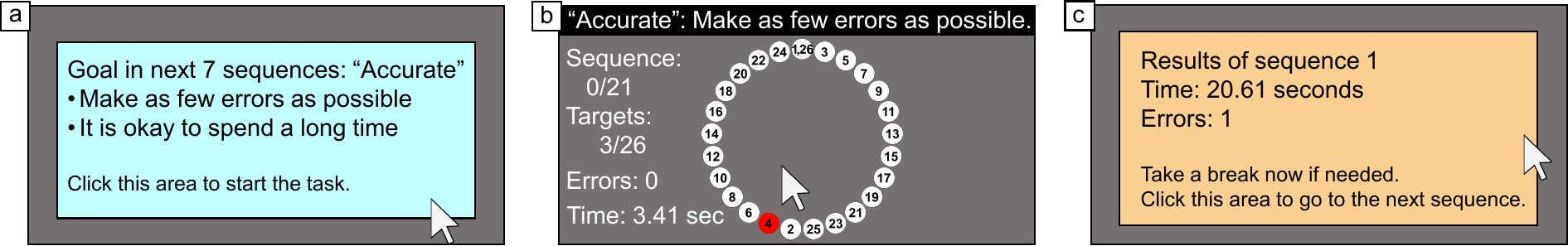}
\caption{Abstract images of the experimental task. (a) Before starting each $\bias$ condition, the goal for the seven sequences including practice was described. (b) Participants clicked the targets in the specified order. The given $\bias$ was always written at the top of the window. (c) After completing a sequence, the results and a message to take a break if needed were displayed.}
\label{fig:studyScreen}
\Description{Images of the crowdsourced experiment. (a) A bias instruction screen before each block; (b) a sequence where participants select 25 red targets in the prescribed order with the current bias shown; (c) post-sequence feedback with time, errors, and a rest prompt.}
\end{figure*}

Our methodology used crowdsourcing via the {\sl Yahoo! Crowdsourcing} platform \cite{YCrowd}.
Crowdsourcing offers advantages for the present study since (i) it is easy to recruit a large sample of participants, and (ii) the fit of Fitts' law models generally improves with larger sample sizes \cite{Yamanaka24ijhcimerit,Yamanaka21hcomp}.

On the other hand, compared to laboratory studies, crowdsourcing poses concerns about experimental control (e.g., devices used) and instruction compliance.
However, performing tasks in each participant's everyday environment can increase ecological validity, and thus these concerns are not necessarily disadvantages.

In addition, previous research reported that crowdsourcing studies can reach the same conclusions as laboratory studies regarding the fit of models predicting $\mt$.
Examples include target pointing \cite{Findlater17,Komarov13,Yamanaka21hcomp,Yamanaka24MobileHCI}, goal crossing \cite{Yamanaka24chiLatency}, scrolling \cite{Schwab19slider}, and path steering \cite{yamanaka23IJHCI,Yamanaka2024jipSteer}.
Moreover, participants follow multiple bias instructions to change their $\mt$s and $\er$s in pointing tasks \cite{Yamanaka22chiBias,Yamanaka24ijhcimerit,Yamanaka24chiLatency}.
These findings support the feasibility of answering our research questions via a crowdsourcing protocol.
If future laboratory-based replications yield conclusions different from ours, they will provide new evidence in the debate over laboratory vs. crowdsourced studies \cite{Findlater17,Komarov13}, but such discussion is outside our scope.

\subsection{Participants}
\label{sec:participants}

We recruited workers with systems running Windows Vista or later and with display resolution at least 1280 $\times$ 850 pixels.
Using the crowdsourcing platform's white list option, we screened out newly-created accounts.
Although the specific criteria were not disclosed, this feature restricted applications to reliable workers based on past participation.

To reduce noise from device diversity, we asked workers to use a mouse if possible.
However, to avoid false claims of mouse use, we stated that any pointing device was allowed.  In the end, we only analyzed data from mouse users.  To increase external validity, we did not control mouse specifications (e.g., DPI, wired/wireless), nor did we require setting the cursor speed or acceleration in the Control Panel.
Previous studies showed that even when display or mouse specifications are unknown for many workers, aggregating mean $\mt$ still yields good model fits \cite{Findlater17,Yamanaka21hcomp,Yamanaka24ijhcimerit}.
We therefore expected similar results here.

We recruited participants with an upper limit of 550 due to budget constraints.
A total of 398 participants completed the task within the one-week posting period, and 52 participants did not use a mouse.
Hereafter, we analyze the remaining 346 mouse users.
Their mean age was 45.6~years ($\sd=$ 11.9).
Each worker received 300~JPY in compensation (about 2~USD as of June 2025).
The average testing time was about 10~min ($\sd=$~3.5).
Thus, the mean effective hourly payment was 1{,}798~JPY (12.4~USD).

\subsection{Apparatus}

The hardware requirements for recruiting participants are given in Section~\ref{sec:participants}.
The software was developed using the \textsf{Hot Soup Processor} programming language, with crowdworkers downloading an executable to perform the task.

The task involved clicking 25 circular targets arranged around a layout circle (Figure \ref{fig:studyScreen}).
The experimental app opened a $1200 \times 800$ pixel window set to refresh at 250~Hz.
The current target appeared in red with others in white.
A sequence of trials comprised 25 consecutive selections under a fixed $A$-$W$ condition.
We required participants to perform according to the currently specified speed-accuracy $\bias$ among $\acc$, $\neu$, and $\fast$.

For each trial, if a click fell inside the target in the first attempt, it was defined as a success; otherwise, it was an error.
In the case of errors, the target flashed yellow and the participant re-aimed for the current target until an on-target click occurred.
However, we logged and analyzed the coordinates and $\mt$ of the first click regardless of success or error.
For example, if the first click missed the target (thus requiring re-aiming), the second click also missed, and the third click finally hit the target, we analyzed the first error click and subsequent re-aiming clicks served only to complete the trial.
By contrast, in ISO 9241-411 and the FittsStudy tool, a first off-target click is logged as an error and the user proceeds to the next target without re-aiming \cite{iso2012,Wobbrock11dim}.
Importantly, in all three cases (our protocol, ISO 9241-411, and FittsStudy), the analysis consistently uses the coordinates of the first click for the current target, whether that click is successful (on-target) or an error (off-target).
We enforced re-aiming to discourage ``racing through the experiment by clicking anywhere'' \cite{Grossman05,Grossman07,Wigdor06,Forlines08,Forlines06,Casiez11Surfpad,Yamanaka22chiBias} and to control movement distance on every trial, which is particularly important in crowdsourced settings where inattentive behavior can degrade data quality and internal validity \cite{Krosnick91satis,Maniaci14satis,Miura16satis}.

\subsection{Procedure}

For each speed-accuracy $\bias$, participants first completed a practice sequence of trials at medium difficulty ($A=460$, $W=50$ pixels; $\id=3.35$ bits) and then completed the 2 amplitudes $\times$ 3 widths = 6 data-collection sequences in random order. 
For each $\bias$ condition, instructions appeared at the top of the window; they remained visible during the task.
The three bias conditions were as follows.
\begin{itemize}
    \item Accurate: Perform the task so as not to make errors as much as possible without worrying about the duration.
    \item Neutral: Perform the task as rapidly and as accurately as possible.
    \item Fast: Perform the task as quickly as possible without worrying about making errors.
\end{itemize}

The order of three $\bias$ conditions was randomized per participant.
After each sequence, we displayed the total time, the number of errors, and a message to rest as needed.
The total number of trials for each participant was 450 ($=$ 3 bias levels $\times$ 2 amplitudes $\times$ 3 widths $\times$ 25 selections per sequence). 

After testing, participants answered a demographic questionnaire, uploaded a log file to our server, and received compensation.
See Appendix~\ref{sec:appendixEthics} for the research ethics approval.

\subsection{Design}
\label{sec:design}

We used a $3 \times 2 \times 3$ within-subjects design.  The independent variables and levels were as follows:

\begin{itemize}
\item Bias (accurate, neutral, fast)
\item Amplitude (320, 500 px)
\item Width (20, 45, 100 px)    
\end{itemize}

With $A$ and $W$ as above, $\id$ ranged from 2.07 to 4.70 bits.
Previous studies with multiple bias conditions used lower and higher upper-end difficulties (e.g., 4.35~bits \cite{Yamanaka24ijhcimerit}, 4.39~bits \cite{Yamanaka24MobileHCI}, and 6.15~bits \cite{Zhai04speed}).
Because high-difficulty tasks are unsuitable for crowdsourcing, due to increasing the risk of dropout, we used lower $\id$s.
Still, our $\id$ range covered difficulties requiring different precision, from ballistic to closed-loop movements \cite{Gan88,Hoffmann16crit}.

The dependent variables were movement time ($\mt$ in seconds), error rate ($\er$ as \% errors) and throughput ($\tp$ in bps).  $\mt$ and $\er$ were recorded directly by the experimental app while $\tp$ was calculated in different ways as per the research questions noted above.

$\mt$ was the time from the correct click on the previous target to the first click in the current trial.
$\er$ was the percentage of trials in a sequence requiring two or more clicks.
Because $\tp$ combines the two independent variables $A$ and $W$ (see Equation~\ref{eqn:defTP}), we did not subject $\tp$ to ANOVA testing.

\section{Results}
\subsection{Data Screening}

We recorded 346 participants $\times$ 450 trials = 155,700 trials.
Following previous studies, we removed trial- and participant-level outliers.
First, we identified spatial outliers whose movement distance for the first click was less than $A/2$ in a trial \cite{Findlater17,MacKenzie08}.
This eliminates obvious errors like double-clicking the previous target.
Although those studies also removed trials whose first click was more than $2W$ from the target center, we did not use this criterion because it is plausible to see many such trials under the $\fast$ condition \cite{Yamanaka24MobileHCI,Yamanaka24ijhcimerit}.

Next, within each participant and $A$-$W$-bias condition, we detected extremely fast or slow trials using the interquartile range ($\mathit{IQR}$) method \cite{Findlater17,Yamanaka21hcomp}.
We removed trials where $\mt$ was more than $3\mathit{IQR}$ below the first quartile or above the third quartile.

At the participant level, we computed each participant's mean $\mt$ across all sequences and removed participants based on the $3\mathit{IQR}$ criterion.
Additionally, we excluded participants for whom 22 or more outliers occurred within any single target condition, because too few data points would remain for computing $\upsigma$.
These criteria excluded four participants.

In total, 3{,}474 trials (2.23\%) were removed at the trial and participant levels.
We analyzed the remaining 152{,}226 trials (97.77\%) from 342 participants.
In crowdsourced Fitts' law studies, both lower (0.5\% \cite{Yamanaka21hcomp}) and higher (4--7\% \cite{Yamanaka24iss,Komarov13,Schwab19slider}) exclusion rates occur; thus, we report no evidence that our data were inappropriate.


\subsection{Movement Time and Error Rate}

\begin{figure*}[t]
\centering
\includegraphics[width=0.8\textwidth]{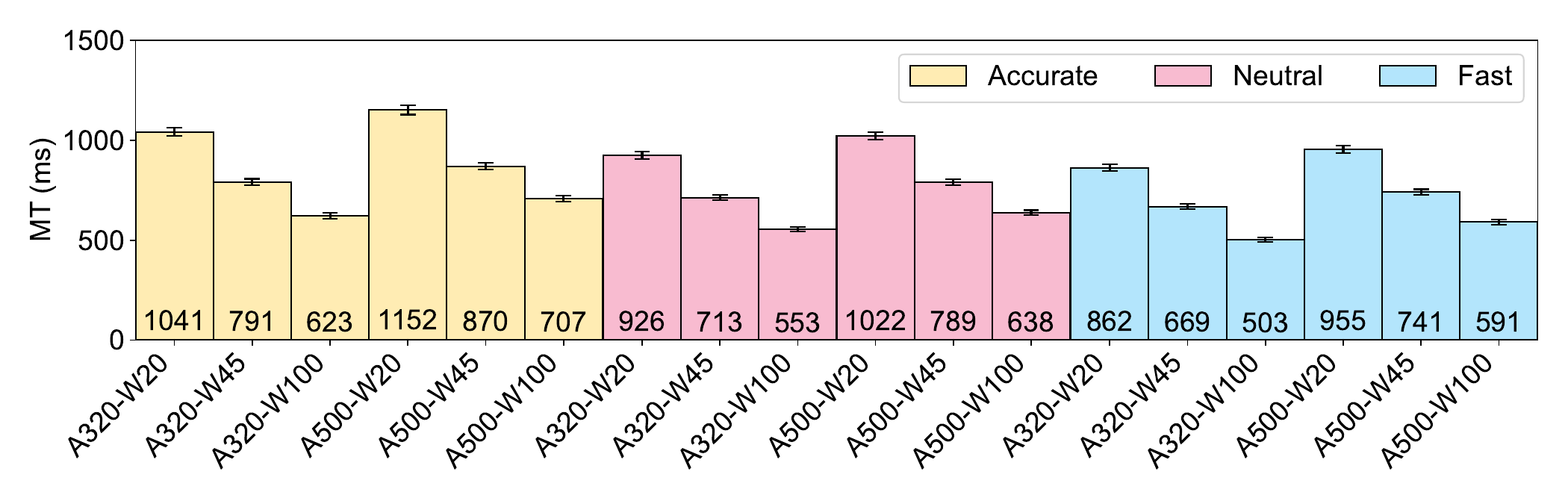}
\caption{Movement time ($\mt$) by amplitude ($A$), width ($W$), and bias.  Throughout this paper, error bars indicate 95\% confidence intervals.}
\label{fig:resBarchartMT}
\Description{Movement time (MT) by amplitude (A), width (W), and bias (Accurate/Neutral/Fast). MT increases with longer A and smaller W, and the three biases are clearly separated; error bars are 95\% CIs.}
\end{figure*}

The grand mean for movement time ($\mt$) was 786~ms per trial.
For the $\acc$, $\neu$, and $\fast$ speed-accuracy bias conditions, the means were 864, 773, and 720~ms, respectively.
Figure \ref{fig:resBarchartMT} shows the results of $\mt$ for each $A$-$W$-$\bias$ condition.

\begin{figure*}[t]
\centering
\includegraphics[width=0.8\textwidth]{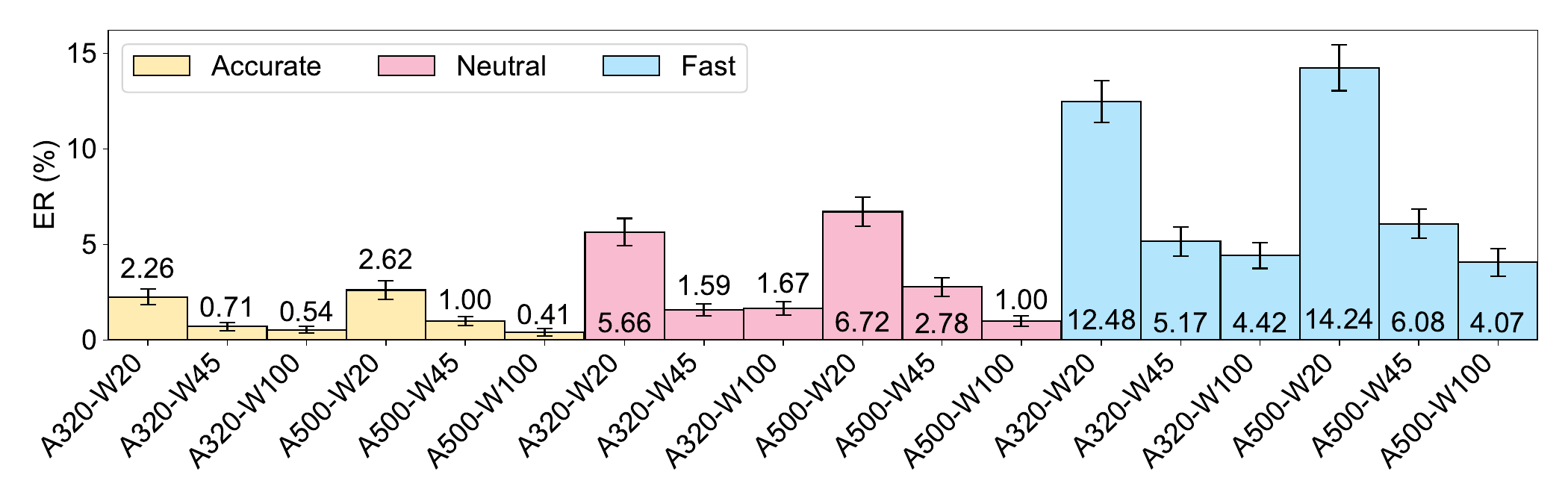}
\caption{Error rate ($\er$) by amplitude ($A$), width ($W$), and bias.}
\label{fig:resBarchartER100}
\Description{Error rate (ER) by A, W, and bias. ER is lowest under Accurate and highest under Fast, with smaller W generally yielding more errors; error bars are 95\% CIs.}
\end{figure*}

The grand mean for error rate ($\er$) was 4.08\%.
By bias condition, the means were 1.26\% ($\acc$), 3.24\% ($\neu$), and 7.74\% ($\fast$).
Figure \ref{fig:resBarchartER100} shows the results of $\er$ for each $A$-$W$-$\bias$ condition.
An Anderson-Darling test \aptLtoX{(\&\#x3B1;$ =.05$)}{($\upalpha =.05$)} showed that the $\mt$ and $\er$ data violated the normality assumption.
We therefore used a non-parametric ANOVA with the aligned-rank transform \cite{Wobbrock11ART,Elkin21ART}.

Tables~\ref{tab:resAnovaMT} and \ref{tab:resAnovaER} show the ANOVA results.
Pairwise tests with the Bonferroni correction showed significant differences ($p<.001$) among all three $\bias$ conditions for both $\mt$ and $\er$ with large effect sizes \cite{Cohen88}.
This indicates that participants sufficiently changed their performance according to the instructions.
Full results of the ANOVA tests are provided in the supplementary materials.

\begin{table}[t]
    \centering
    \caption{ANOVA results for $\mt$}
    \label{tab:resAnovaMT}
    \begin{tabular}{l c c c}
      \hline
      Factor                   & $F$                         & $p$      & $\upeta_\mathrm{p}^2$ \\ 
      \hline
      $\Bias$                  & $F_{2,682}=473.7$      & $<.001$ & .5815     \\
      $A$                      & $F_{1,341}=2261$     & $<.001$ & .8690     \\
      $W$                      & $F_{2,682}=7871$     & $<.001$ & .9585     \\
      $\Bias \times A$         & $F_{2,682}=3.450$        & $<.05$ & .0100     \\
      $\Bias \times W$         & $F_{4,1364}=97.75$      & $<.001$ & .2228     \\
      $A \times W$             & $F_{2,682}=37.28$       & $<.001$ & .0986     \\
      $\Bias \times A \times W$& $F_{4,1364}=2.287$       & $.058$ & .0067     \\
      \hline
    \end{tabular}
    \Description{Aligned-rank transform ANOVA for movement time (MT). Bias, A, and W have large, significant main effects; several interactions are also significant, confirming that instructions and task geometry modulate MT.}
\end{table}

\begin{table}[t]
    \centering
    \caption{ANOVA results for $\er$}
    \label{tab:resAnovaER}
    \begin{tabular}{l c c c}
      \hline
      Factor                   & $F$                         & $p$      & $\upeta_\mathrm{p}^2$ \\ 
      \hline
      $\Bias$                  & $F_{2,682}=799.9$      & $<.001$ & .7011     \\
      $A$                      & $F_{1,341}=77.85$       & $<.001$ & .1859     \\
      $W$                      & $F_{2,682}=788.9$      & $<.001$ & .6982     \\
      $\Bias \times A$         & $F_{2,682}=38.36$       & $<.001$ & .1011     \\
      $\Bias \times W$         & $F_{4,1364}=241.3$     & $<.001$ & .4144     \\
      $A \times W$             & $F_{2,682}=48.30$       & $<.001$ & .1241     \\
      $\Bias \times A \times W$& $F_{4,1364}=13.34$      & $<.001$ & .0376     \\
      \hline
    \end{tabular}
    \Description{Aligned-rank transform ANOVA for error rate (ER). Bias and W show large main effects, with additional significant interactions, indicating systematic accuracy changes across instructions and target geometry.}
\end{table}

\subsection{Visualization and Normality of Click Coordinates}

\begin{figure*}[t]
\centering
\includegraphics[width=1.0\textwidth]{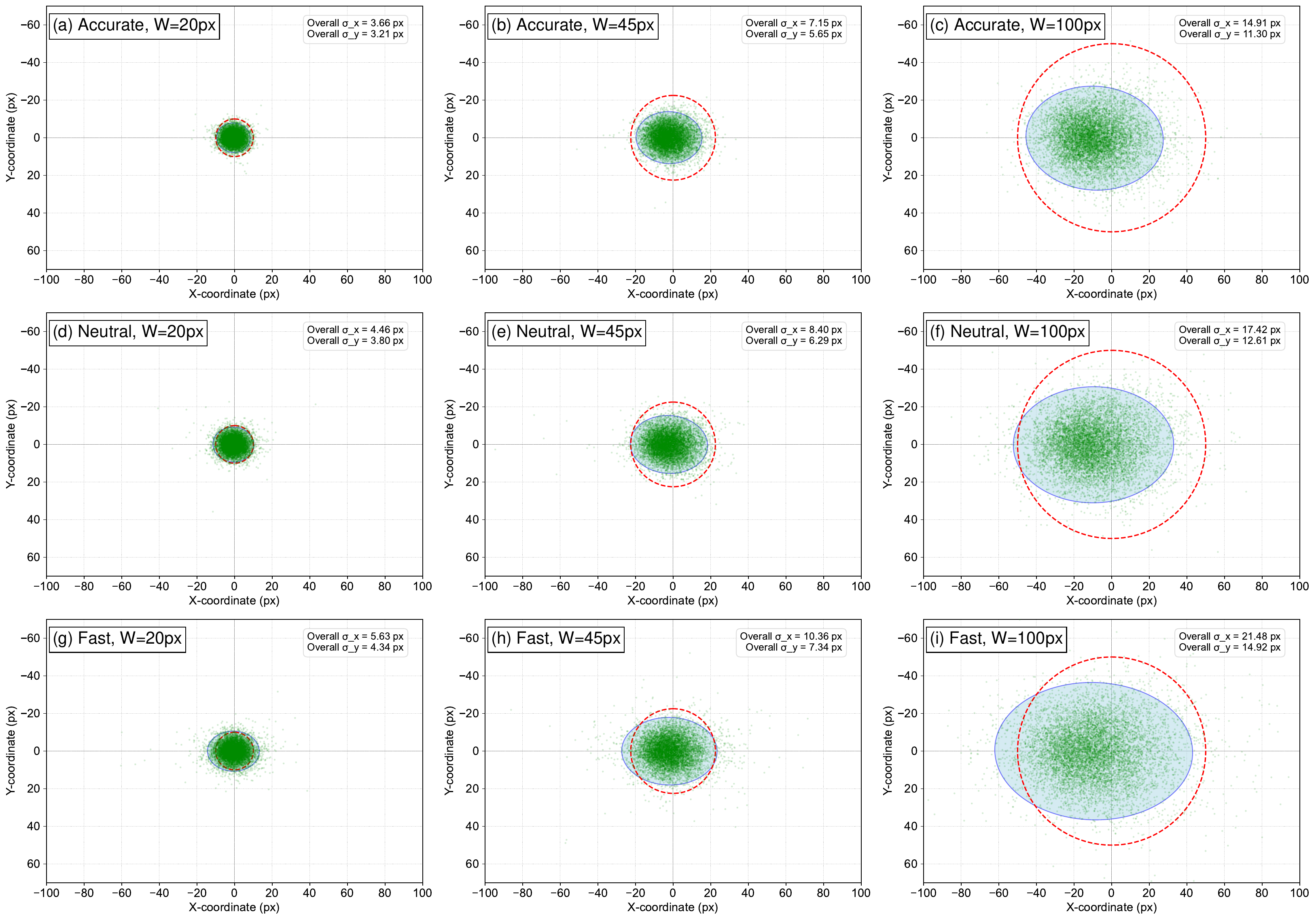}
\caption{Endpoint distributions for $A=320$ pixels. Green dots are endpoints, red dashed circles are targets, and blue ellipses are 95\% confidence ellipses. The task axis is rotated so that the direction from the previous to the current target center is positive $x$.}
\label{fig:resMergedNineEndpoints}
\Description{Endpoint distributions for A=320 px across biases and widths after rotating to the TT axis. Green points (clicks), dashed red targets, and blue 95\% confidence ellipses show that spread increases with speed‑oriented bias and with smaller targets.}
\end{figure*}

To visually verify that participants altered behavior across $\bias$ conditions, we plotted the endpoint spreads.  See Figure~\ref{fig:resMergedNineEndpoints}.
Using the TT task axis, we rotated each trial so that the $x$-axis pointed to the right ($\uptheta=0^\circ$).

We tested 1D normality of endpoint distributions along the $x$-axis using Shapiro-Wilk test and 2D normality using the Henze-Zirkler test \aptLtoX{(\&\#x3B1; $=.05$)}{($\upalpha =.05$)}.
For each bias condition -- $\acc$, $\neu$, and $\fast$ -- we conducted $2_A \times 3_W \times 342_\mathrm{workers}=2{,}052$ tests.
In 1D, 1{,}827 (89.0\%), 1{,}813 (88.4\%), and 1{,}766 (86.1\%) were normal, respectively.
In 2D, 1{,}869 (91.1\%), 1{,}862 (90.7\%), and 1{,}841 (89.7\%) were normal.
For touch-based studies with circular targets, 2D normality was reported at 68.3--76.7\% \cite{Yamanaka20issFFF} or 100\% \cite{Bi13a}.  Our results fall inside this range.

As expected, to avoid errors, participants were more careful for smaller $W$, resulting in smaller endpoint variability.
$\upsigma_\mathit{x}$ and $\upsigma_\mathit{y}$ also increased as instructions emphasized speed.
Together with the $\mt$ and $\er$ results, these visualizations verify that participants followed the $\bias$ instructions.

\subsection{Model Fit}

Here, model-fit and $\tp$-stability values are reported with many digits due to the small differences between the TT and CT axis tests.
The compared models are summarized in Table~\ref{tab:comparedModels}.

\begin{figure*}[t]
\centering
\includegraphics[width=1.0\textwidth]{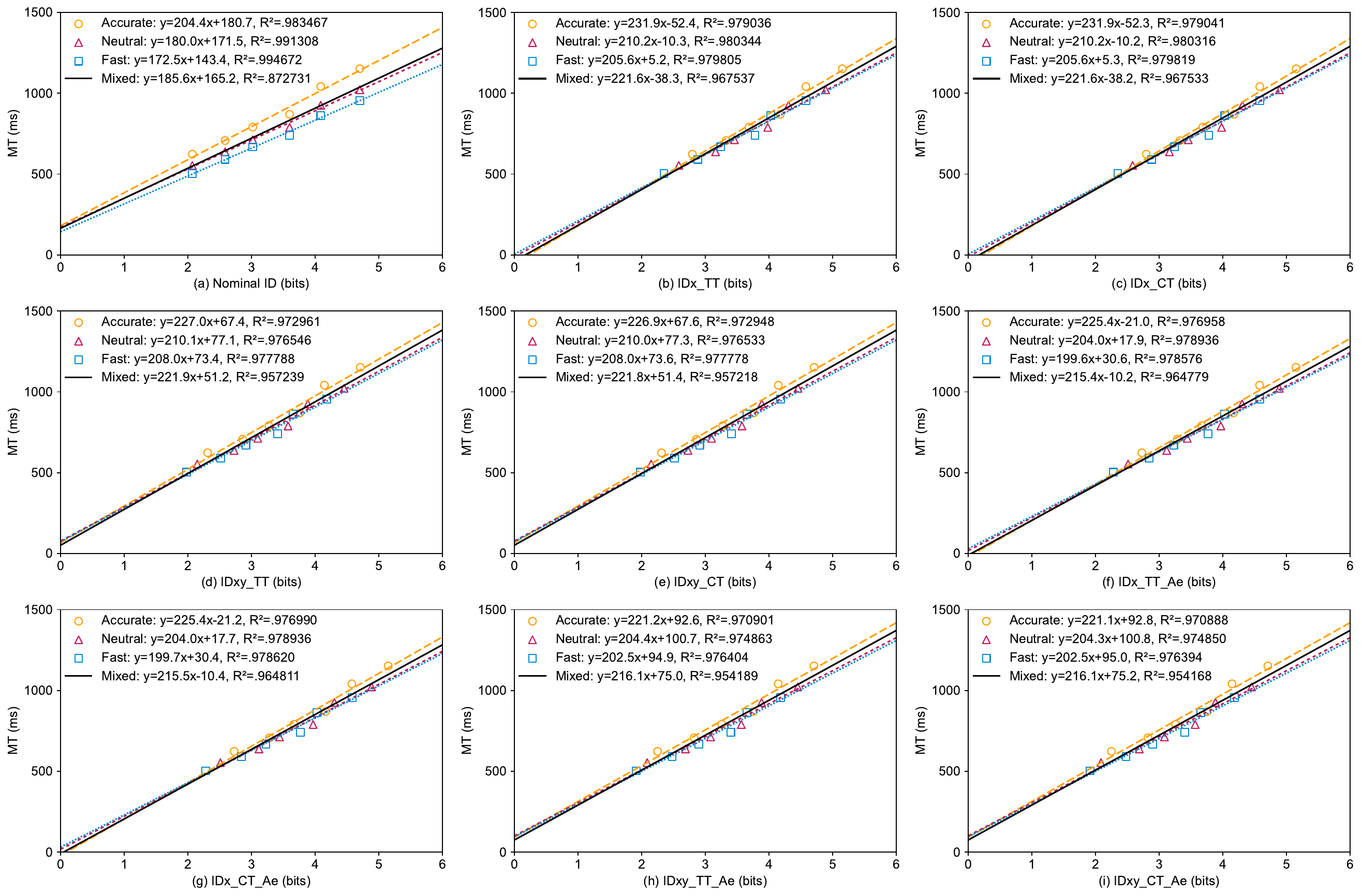}
\caption{Fitts' law regression models with nine $\id$ computation methods. Using nominal $\id$ yielded the lowest $R^2$ of $.872731$ for the mixed-bias $\mt$ data, while the other eight effective-width models showed $R^2 > .95$.}
\label{fig:resRegressNine}
\Description{Fitts' law regressions comparing nine ID computation methods. Per-bias fits remain strong in both, but in the mixed (all-bias) analysis the effective-width model aligns points more closely to a single line, increasing R^2.}
\end{figure*}

We replicated two findings from previous studies.
(1) When analyzing $\acc$, $\neu$, and $\fast$ separately and regressing six data points ($=2_A\times3_W$), nominal $\id$ yielded better fit than $\ide$ for each method of computing $\upsigma$.
Specifically, using $\id$ yielded $R^2=$ .983467, .991308, and .994672 (Figure~\ref{fig:resRegressNine}a), whereas one $\ide$ model, $\id_\xTT$, yielded $R^2$ lower by .004 to .015 points (Figure~\ref{fig:resRegressNine}b).
(2) In the mixed condition regressing 18 data points across the three biases, $\id$ yielded $R^2=.872731$ (Figure~\ref{fig:resRegressNine}a), whereas the best $\ide$ model, $\id_\xTT$, achieved $R^2=.967537$ (Figure~\ref{fig:resRegressNine}b).
These results show that $W_\mathrm{e}$ improves the model fit by normalizing bias effects \cite{Zhai04speed,Yamanaka24ijhcimerit,Yamanaka24MobileHCI}.
This finding also applied to other $\upsigma$ models (Figure~\ref{fig:resRegressNine}c--i).

\begin{table*}[t]
\centering
\caption{Models compared for Fitts' law fitting and $\tp$ calculation}
\label{tab:comparedModels}
\begin{tabular}{ll}
\hline
\textbf{Model} & \textbf{Description} \\
\hline
Nominal $\id$ & Baseline Fitts' law model using nominal width and nominal amplitude \\
\rowcolor{gray!10}\hline
$\id_\xTT$ & Univariate $\upsigma_\mathit{x}$ effective width, TT task axis, nominal amplitude \\
$\id_\xCT$ & Univariate $\upsigma_\mathit{x}$ effective width, CT task axis, nominal amplitude \\
\rowcolor{gray!10}
$\id_\xyTT$ & Bivariate $\upsigma_\mathit{xy}$ effective width, TT task axis, nominal amplitude \\
$\id_\xyCT$ & Bivariate $\upsigma_\mathit{xy}$ effective width, CT task axis, nominal amplitude \\
\hline
\rowcolor{gray!10}
$\id_\xTTAe$ & Univariate $\upsigma_\mathit{x}$ effective width, TT task axis, effective amplitude \\
$\id_\xCTAe$ & Univariate $\upsigma_\mathit{x}$ effective width, CT task axis, effective amplitude \\
\rowcolor{gray!10}
$\id_\xyTTAe$ & Bivariate $\upsigma_\mathit{xy}$ effective width, TT task axis, effective amplitude \\
$\id_\xyCTAe$ & Bivariate $\upsigma_\mathit{xy}$ effective width, CT task axis, effective amplitude \\
\hline
\end{tabular}
\Description{Models compared for Fitts' law and TP calculations. The table defines nominal ID and eight effective-width variants, indicating sigma type (univariate/bivariate), task‑axis definition (TT/CT), and whether nominal A or effective Ae is used.}
\end{table*}

\begin{figure*}[t]
\centering
\includegraphics[width=1.0\textwidth]{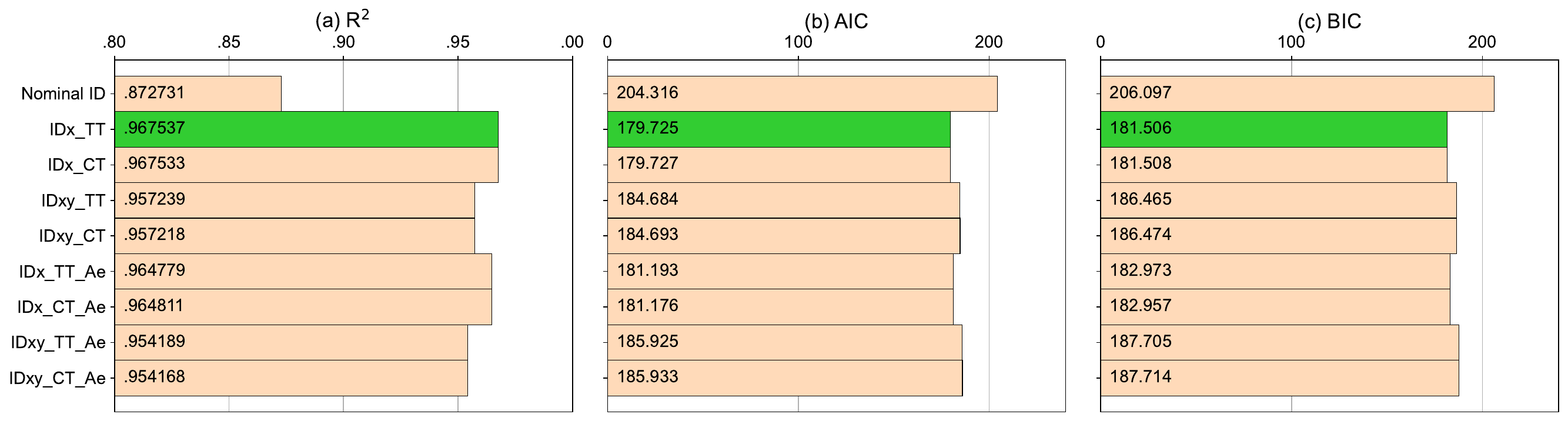}
\caption{Model-fit metrics in the Mixed condition. Throughout this paper, green bars indicate the best result among the nine candidate models, i.e., highest $R^2$ and lowest AIC and BIC.}
\label{fig:resFitMixed}
\Description{Model-fit metrics (Mixed condition) across ID variants. Univariate sigma_x with the TT axis (IDxTT) yields the highest R^2 and the lowest AIC/BIC, while bivariate sigma_xy variants fit worse.}
\end{figure*}

More importantly, we evaluated which $\upsigma$ computation yields the highest fit in the mixed condition.
As shown in Figure~\ref{fig:resFitMixed}a, $\id_\xTT$ yielded the highest $R^2=.967537$.
The next best, $\id_\xCT$, had $R^2=.967533$, an almost negligible difference.
This is reasonable given the small definitional difference between TT and CT task axes.
Similarly, the $R^2$ difference from the two univariate models using $A_\mathrm{e}$ ($\id_\xTTAe$ and $\id_\xCTAe$) was below .003.

In contrast, models using bivariate standard deviation had $R^2<.96$.
To analyze differences in fit statistically, we used AIC (Akaike information criterion) and BIC (Bayesian information criterion).
Lower AIC and BIC indicate more explanatory models.
The following rules of thumb apply for AIC: (1) If differences from the best model are less than 2, there is evidence to support the candidate models. (2) Differences of 2--4 indicate considerable support. (3) Differences of 4--7 indicate much less support. (4) Differences $>$10 indicate essentially no support \cite{Burnham2003}.
For BIC, differences of 0--2 are not significant, 2--6 positive, 6--10 strong, and $>$10 very strong \cite{Robert95bic}.

By these criteria, Figure~\ref{fig:resFitMixed}b and Figure~\ref{fig:resFitMixed}c show that the best model $\id_\xTT$ had AIC = 179.725, whereas the four bivariate models had AIC $\ge$ 184.684.
Thus, the AIC differences exceeded 4, indicating much less support for the bivariate models.
Similarly, the BIC for $\id_\xTT$ differed from those for the bivariate models by more than 4, providing positive support for $\id_\xTT$.
Among the four univariate models, there were no significant differences in AIC and BIC ($<$2).

\subsection{Throughput and its Stability}

\begin{figure*}[t]
\centering
\includegraphics[width=1.0\textwidth]{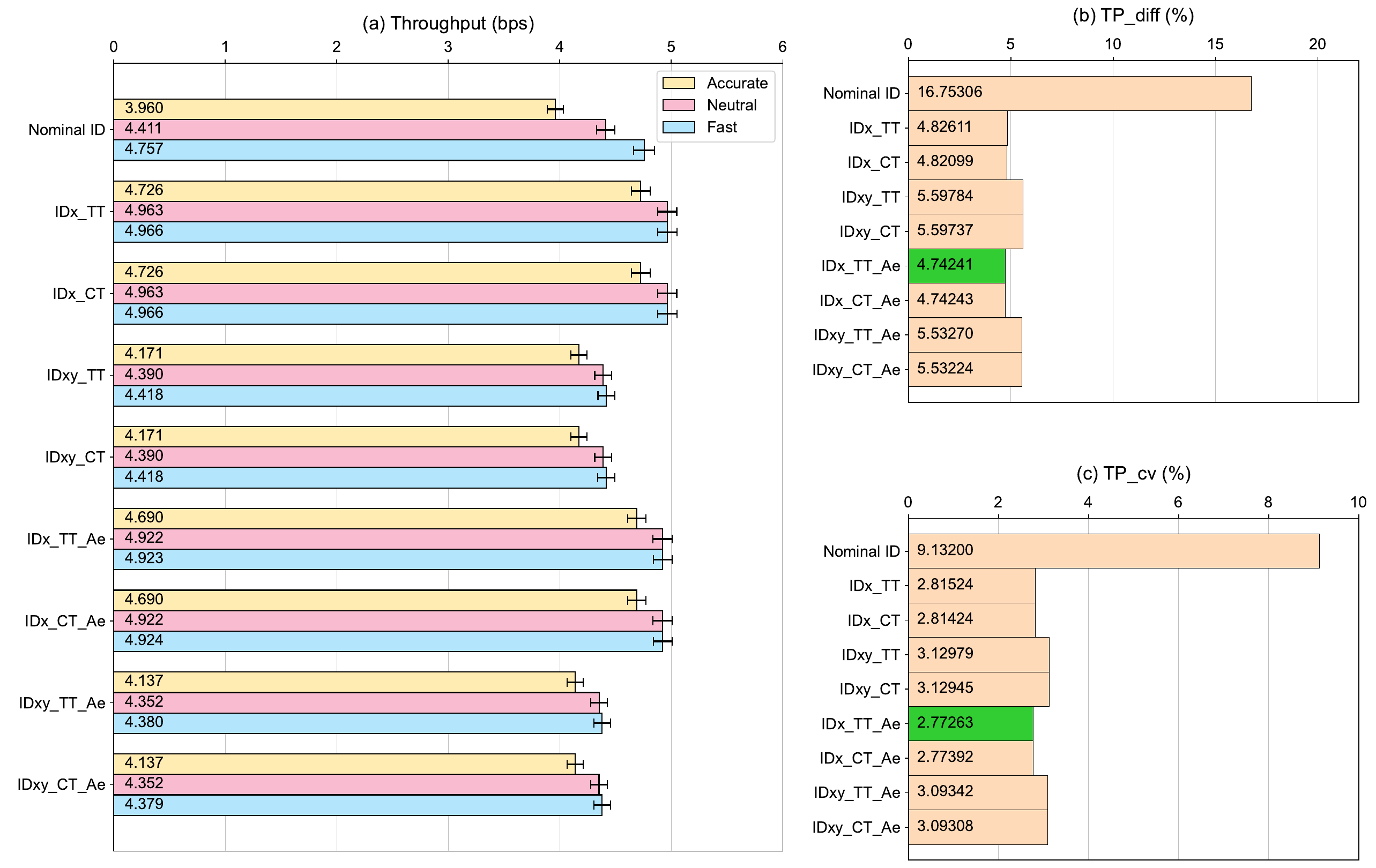}
\caption{$\tp$ values and stability metrics.}
\label{fig:resTP}
\Description{Throughput (TP) by bias and TP‑stability metrics across ID variants. Univariate models show smaller spread of TP across biases, with IDxTTAe slightly improving stability relative to IDxTT.}
\end{figure*}

Using the ISO-recommended $\tp$ calculation method of $\id_\xTT$, the grand mean for throughput ($\tp$) was 4.885~bps.
For the $\acc$, $\neu$, and $\fast$ speed-accuracy bias conditions, the means were 4.726, 4.963, and 4.966~bps, respectively.
However, Figure~\ref{fig:resTP}a clearly shows that $\tp$s varied depending on the $\id$ calculation.
We thus analyze its stability across the three $\bias$ conditions.

$\tp$ stability has a few possible interpretations.  Yamanaka et al. used $\tp_\mathrm{diff}$ defined as the ratio of maximum and minimum $\tp$ across the three biases: $\tp_\mathrm{diff} = 100\% \times (\tp_\mathrm{max} - \tp_\mathrm{min})/\tp_\mathrm{max}$ \cite{Yamanaka22chiBias,Yamanaka24MobileHCI}.
Zhang et al. used $\tp_\mathrm{cv}$ defined as the coefficient of variation of $\tp$s across the three biases: $\tp_\mathrm{cv} = 100\% \times \sd / \mathit{Mean}$ \cite{Zhang19text}.
Lower values indicate better bias normalization for both metrics.

Figure~\ref{fig:resTP}b--c shows the results of two stability metrics.
$\id_\xTTAe$ had the lowest (best) values on both metrics.
Compared to the best-fit model $\id_\xTT$ in terms of $R^2$, the differences were $4.82611 - 4.74241 = 0.08370$ points for $\tp_\mathrm{diff}$ and $2.81524 - 2.77263 = 0.04261$ points for $\tp_\mathrm{cv}$.\footnote{We follow APA guidelines. Because $R^2$ and its differences lie between 0 and 1, we omit the leading zero before the decimal point. By contrast, $\diff$ and $\cv$ in \% and their differences in points can take values $\ge$ 1, so we include the integer digit.}
In contrast, the four bivariate models had $\tp_\mathrm{diff}\ge 5.5\%$ and $\tp_\mathrm{cv}\ge 3.0\%$, which were less desirable than univariate models.

\subsection{Summary}

In the mixed condition, $\id_\xTT$ achieved the highest model fit.
Based on AIC and BIC, we found no significant differences among the other univariate models.
The four bivariate models had $R^2$ at least .01 points lower, and their AIC and BIC provided much less support relative to the best model.
We therefore recommend adopting univariate models to normalize speed-accuracy bias.

Among univariate models, those using the TT task axis ($\id_\xTT$ and $\id_\xTTAe$) had slightly higher fit and lower $\tp$ variability than the other two.
$\id_\xTT$ had the highest fit, and $\id_\xTTAe$ had the lowest variability, with only minor differences.
There is thus no clear evidence to reject either.

\section{Simulation}

\begin{figure*}[t]
\centering
\includegraphics[width=1\textwidth]{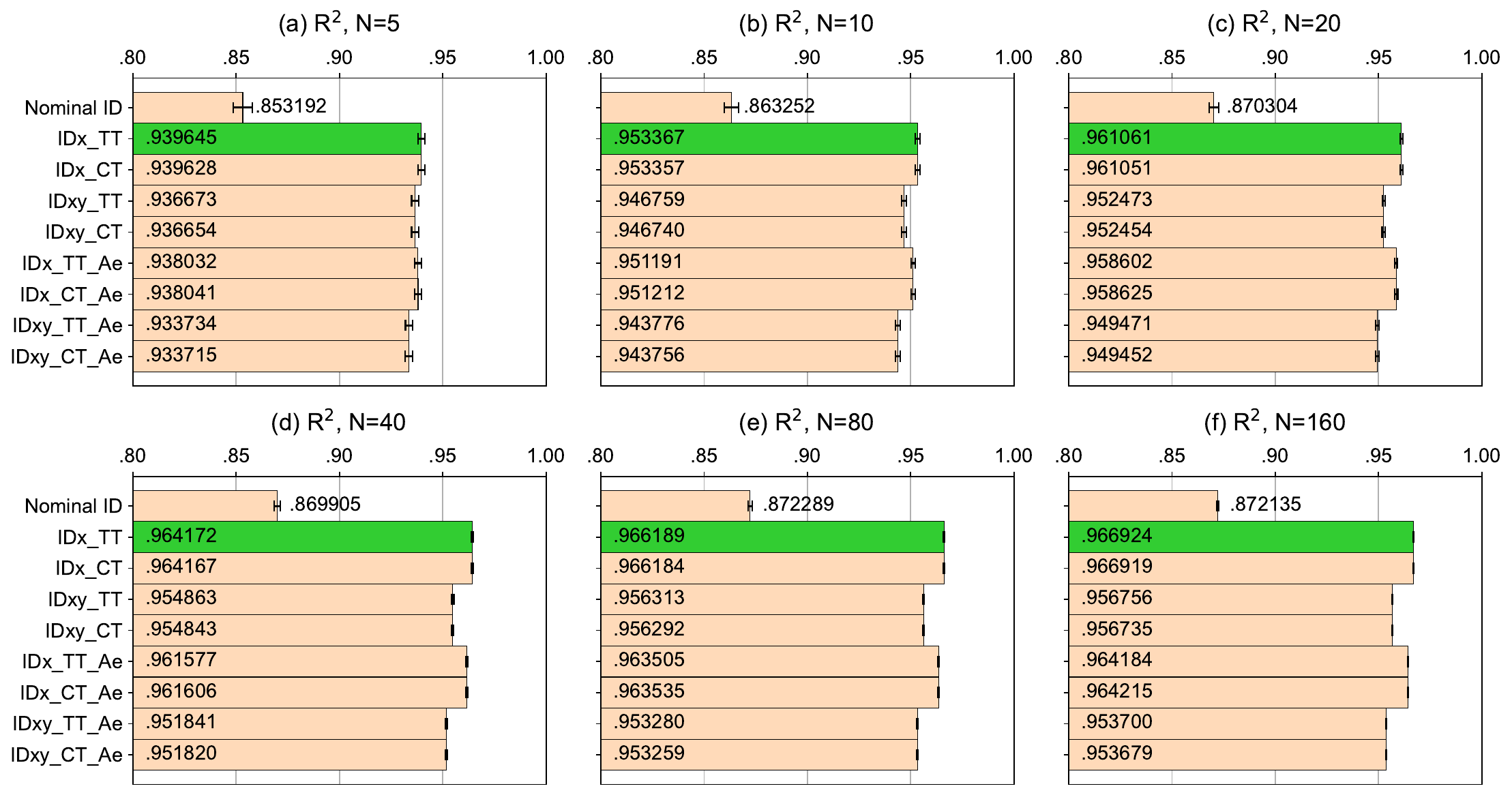}
\caption{Average $R^2$ from the simulation.}
\label{fig:simFitR2}
\Description{Simulation; average R^2 versus sample size N from Monte Carlo subsets. IDxTT consistently attains the highest average R^2 across N.}
\end{figure*}

\begin{figure*}[t]
\centering
\includegraphics[width=1\textwidth]{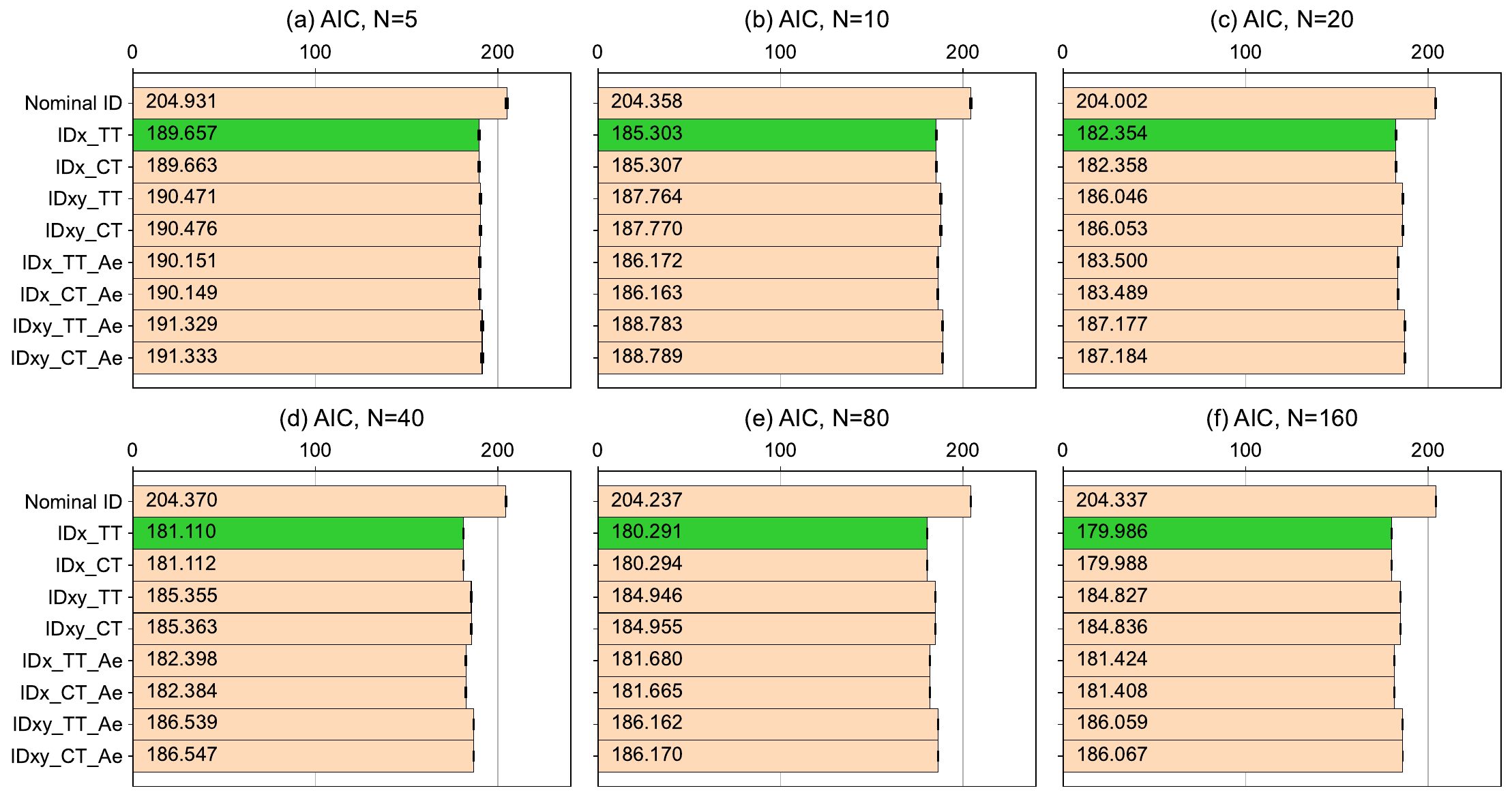}
\caption{Average $\aic$ from the simulation.}
\label{fig:simFitAIC}
\Description{Simulation; average AIC versus sample size N. The univariate sigma_x models achieve lower (better) AIC than sigma_xy models across all N.}
\end{figure*}

\begin{figure*}[t]
\centering
\includegraphics[width=1\textwidth]{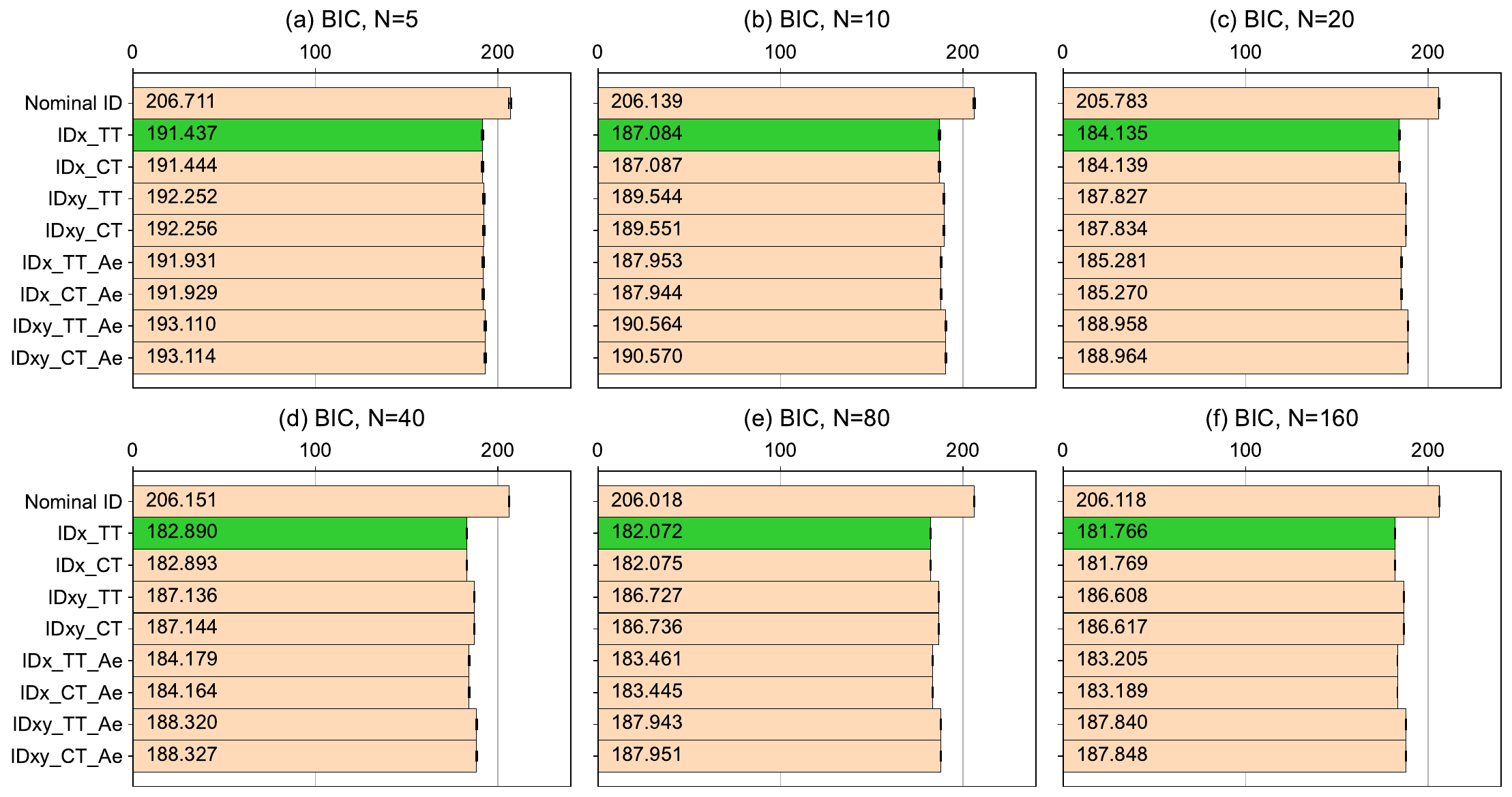}
\caption{Average $\bic$ from the simulation.}
\label{fig:simFitBIC}
\Description{Simulation; average BIC versus sample size N. Patterns mirror AIC: univariate sigma_x models dominate, indicating stronger support than sigma_xy models.}
\end{figure*}

\begin{figure*}[t]
\centering
\includegraphics[width=1\textwidth]{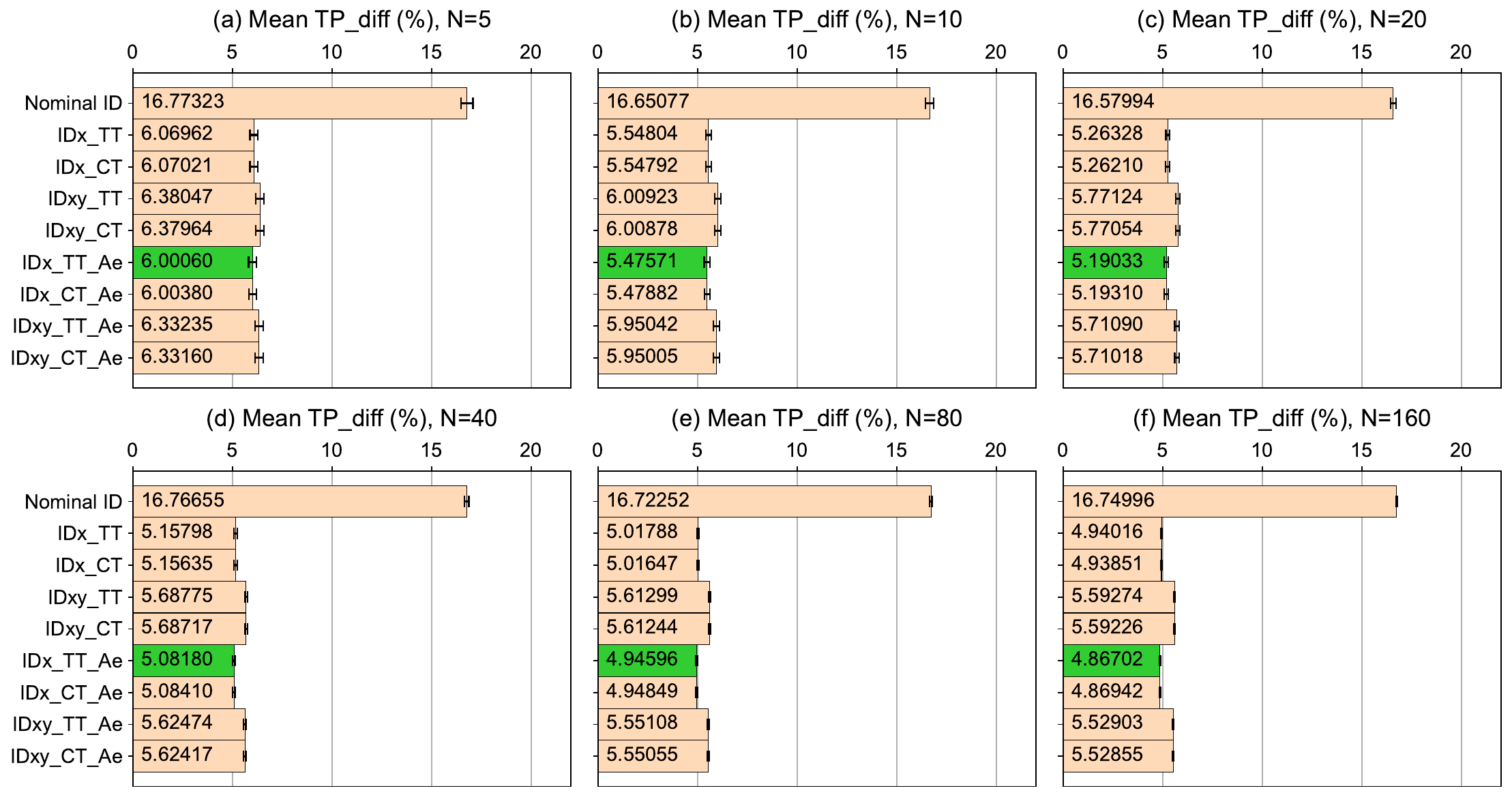}
\caption{Average $\tp_\mathrm{diff}$ from the simulation.}
\label{fig:simTPdiff}
\Description{Simulation; average TPdiff (percent spread of TP across biases) versus N. IDxTTAe yields the lowest TPdiff (best stability) across sample sizes, closely followed by IDxTT.}
\end{figure*}

\begin{figure*}[t]
\centering
\includegraphics[width=1\textwidth]{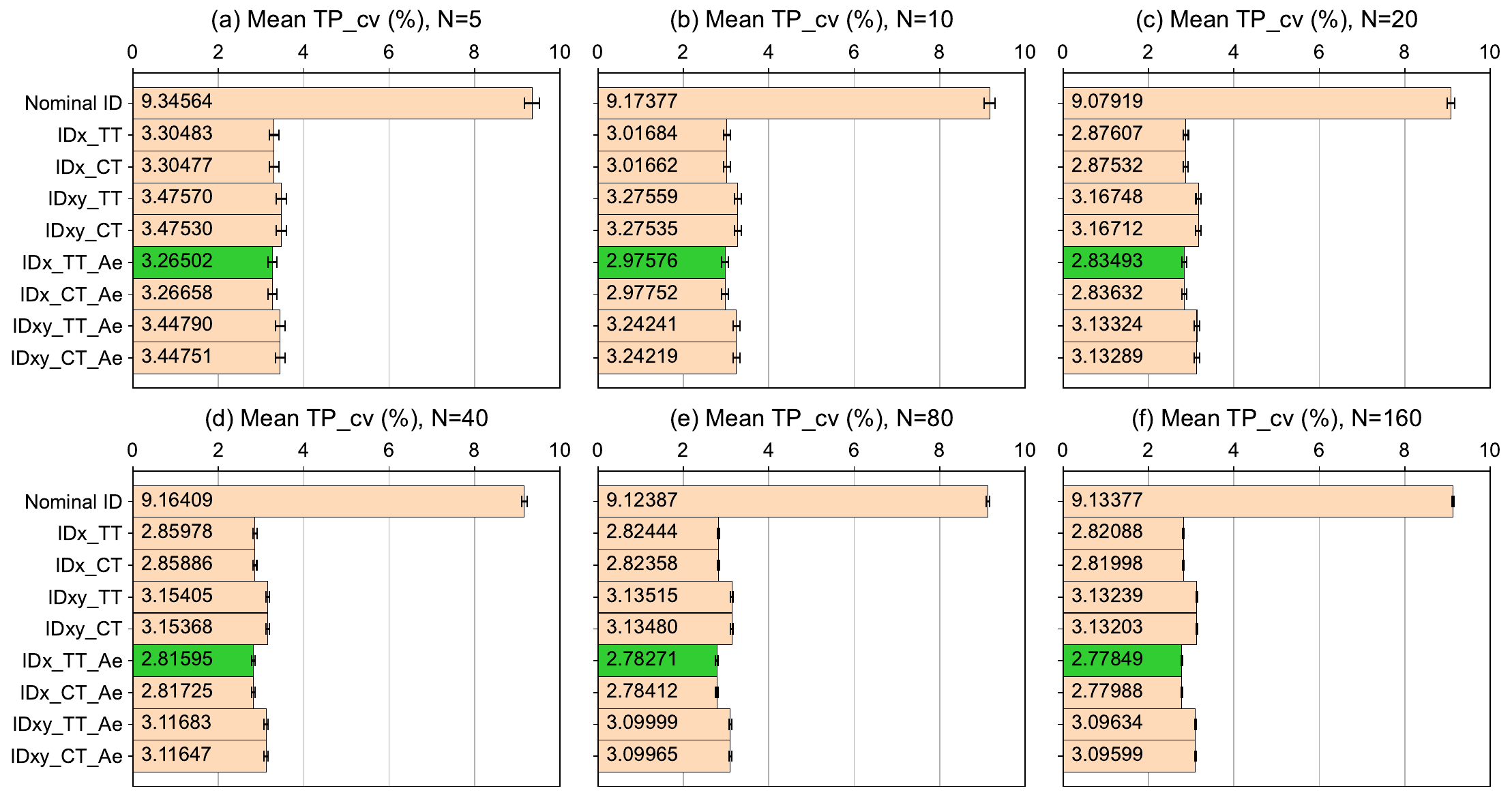}
\caption{Average $\tp_\mathrm{cv}$ from the simulation.}
\label{fig:simTPcv}
\Description{Simulation; average TPcv (coefficient of variation of TP across biases) versus N. Stability improves with N overall, with IDxTTAe most stable and univariate models outperforming bivariate models.}
\end{figure*}

\begin{figure*}[t]
\centering
\includegraphics[width=1.0\textwidth]{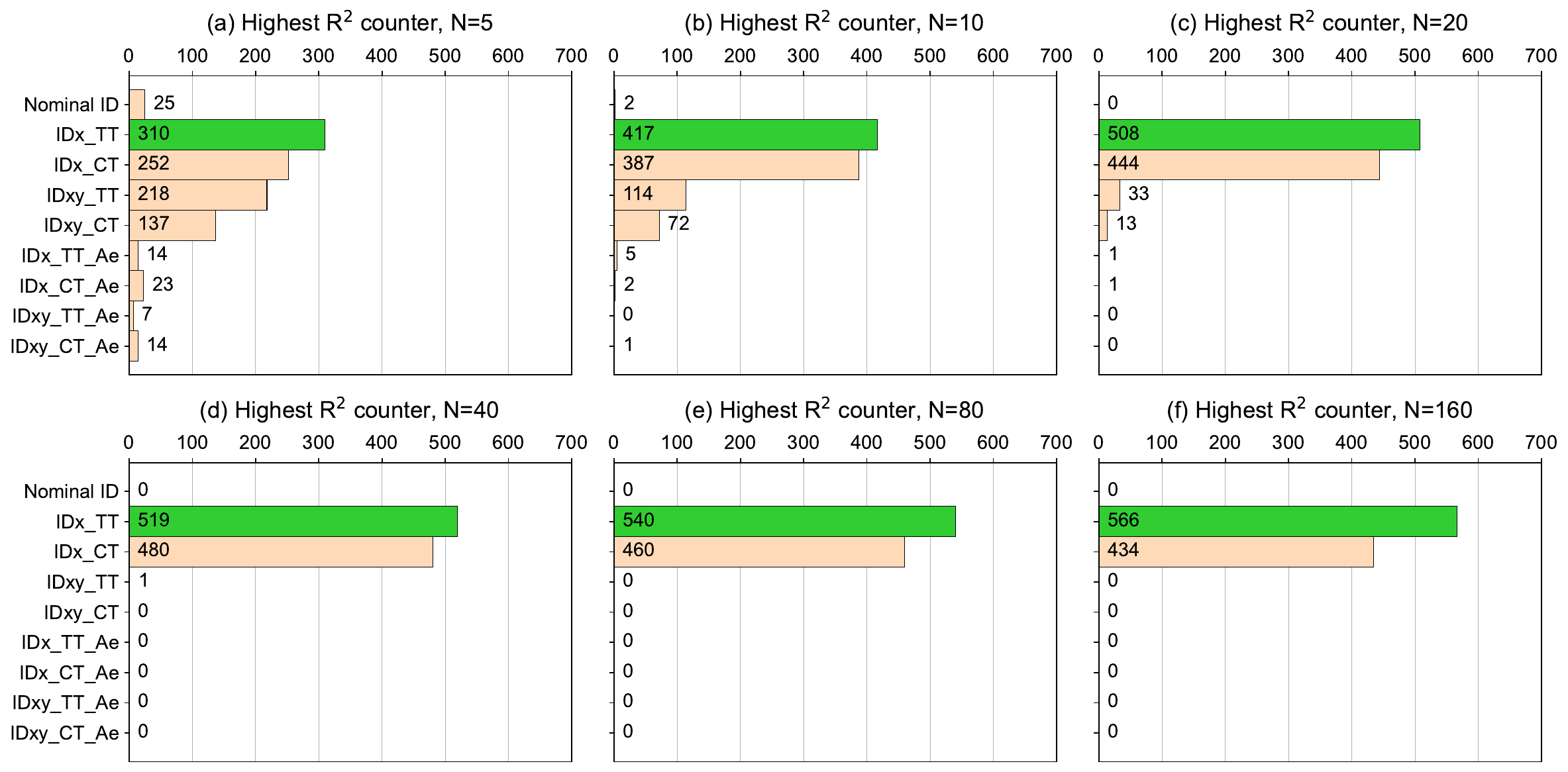}
\caption{Counts of iterations (out of 1{,}000) where each model achieved the highest $R^2$.}
\label{fig:simCounterFit_R2}
\Description{Simulation; counts of runs (out of 1,000) where each model achieved the best R^2 at each N. IDxTT wins most frequently, and the chance that a bivariate model wins drops rapidly as N increases.}
\end{figure*}

\begin{figure*}[t]
\centering
\includegraphics[width=1.0\textwidth]{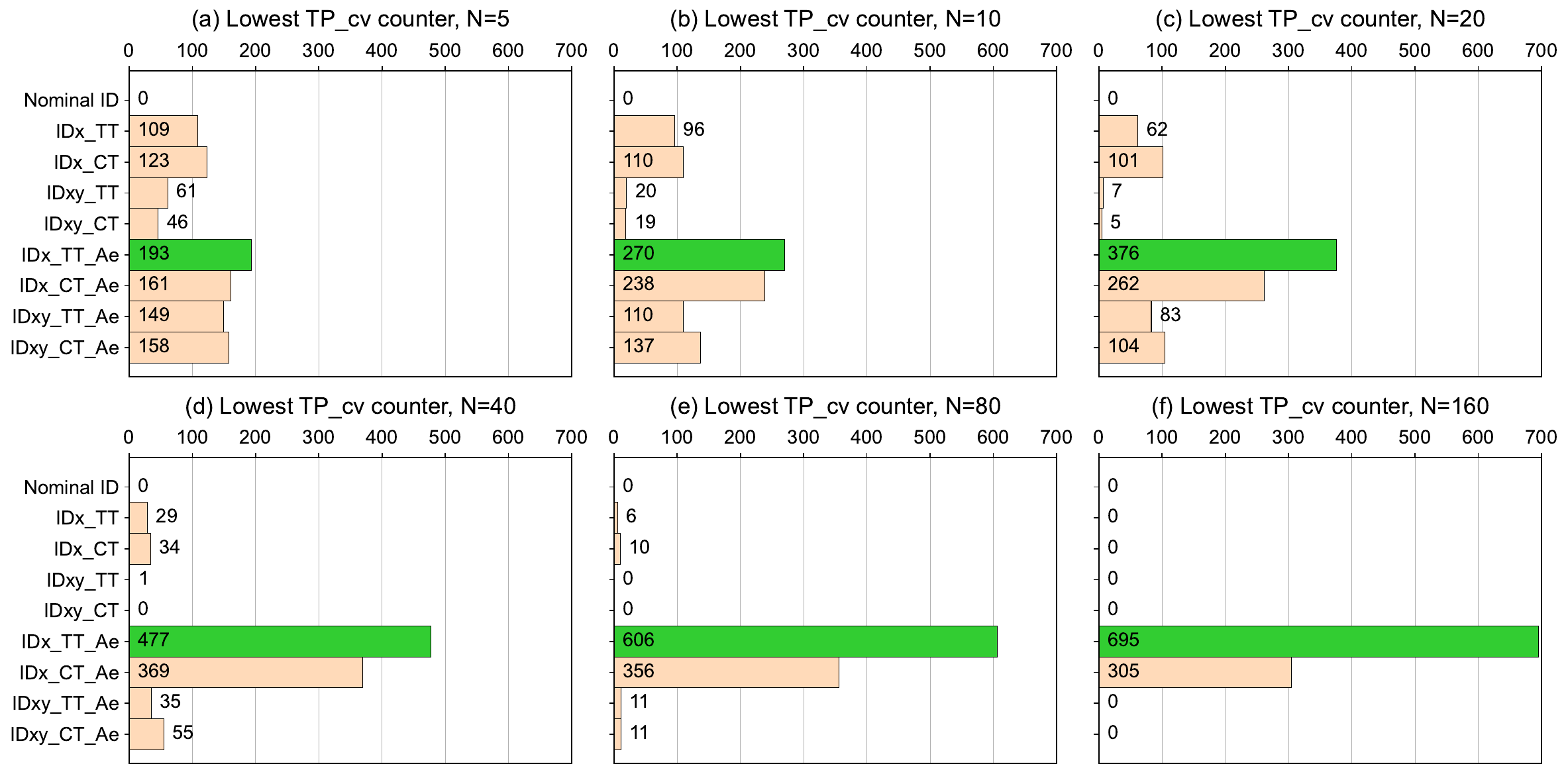}
\caption{Counts of iterations (out of 1{,}000) where each model achieved the lowest $\tp_\mathrm{cv}$.}
\label{fig:simCounterTPavg_CV}
\Description{Simulation; counts of runs where each model achieved the best TPcv at each N. IDxTTAe most often yields the lowest variability; bivariate models occasionally win at very small N but lose competitiveness as N grows.}
\end{figure*}

We showed that $\id_\xTT$ and $\id_\xTTAe$ best normalized the speed-accuracy bias for a large sample of 342 participants.
In practice, however, researchers using ISO-style tasks to compare devices, interaction techniques, or user groups rarely recruit so many participants.
In our survey (Table~\ref{tab:studies_sigma_xy}), sample sizes ranged from 6 to 52, with a mean of 21.5.
Researchers may thus want a model that appropriately normalizes bias even for relatively small samples.

We therefore proceeded with a simulation using randomly sampled subsets of the 342 participants tested.  The goal was to test whether a different model would perform better if only those participants were recruited.
Ideally, as in our main analysis, we would identify participant-level outliers within the sampled set (e.g., for $N=20$), but doing so would entangle our focus of the sample-size effect with model selection.
We thus sampled from the 342 participants after removing the four participant-level outliers identified earlier.

\subsection{Simulation Procedure}

To cover a wide range of sample sizes smaller and larger than typical studies, we set $N$ to 5, 10, 20, 40, 80, and 160.
From the 342 participants, we sampled $N$ participants and computed the mixed-condition model-fit metrics ($R^2$, AIC, BIC) and $\tp$-stability metrics ($\tp_\mathrm{diff}$ and $\tp_\mathrm{cv}$) as in the main analysis.
To handle randomness, we iterated this process 1{,}000 times for each $N$.

We averaged the metrics across iterations per model.
We also counted, for each metric, how many times each model was the best in the 1{,}000 iterations.
This Monte Carlo procedure estimates the probability that a model is best.
For example, if $\id_\xyTTAe$ achieved the highest $R^2$ in 8 of 1{,}000 iterations, the probability is $p=.008<.01$.

\subsection{Simulation Results}

Figures~\ref{fig:simFitR2}, \ref{fig:simFitAIC}, and \ref{fig:simFitBIC} show average model-fit metrics and Figures~\ref{fig:simTPdiff} and \ref{fig:simTPcv} show average $\tp$-stability metrics.
Overall, results matched those using all 342 participants regardless of $N$.
That is, $\id_\xTT$ was best for the three fit metrics, and $\id_\xTTAe$ was best for the two stability metrics.

We further show, in Figures~\ref{fig:simCounterFit_R2} and \ref{fig:simCounterTPavg_CV}, the counts of how often each model was best for fit and stability, respectively.
Because $R^2$, AIC, and BIC aligned (the highest $R^2$ yields the lowest AIC and BIC), and $\tp_\mathrm{diff}$ results mirrored $\tp_\mathrm{cv}$, we report $R^2$ and $\tp_\mathrm{cv}$ as representatives.

For $N=5$ (Figure~\ref{fig:simCounterFit_R2}a), $\id_\xTT$ most frequently achieved the highest $R^2$, as in our main analyses.
Bivariate models also had a nontrivial chance of being best: $218+137+7+14=376$ times in total.
However, this decreases as $N$ increases.
At $N=20$ (Figure~\ref{fig:simCounterFit_R2}c), the bivariate models were best $33+13+0+0=46$ times in total, $p=.046<.05$.
At $N=80$, they had no chance (Figure~\ref{fig:simCounterFit_R2}e).

Similarly, for $\tp_\mathrm{cv}$ at $N=5$ (Figure~\ref{fig:simCounterTPavg_CV}a), $\id_\xTTAe$ most frequently had the lowest value.
Even as $N$ increased, bivariate models remained competitive, with $7+5+83+104=199$ wins at $N=20$.
Ultimately, at $N=160$, bivariate models had no wins.

These results indicate that the safest choices are $\id_\xTT$ for model fit and $\id_\xTTAe$ for $\tp$ stability when researchers cannot recruit large samples.
With small $N$, bivariate models occasionally prevailed, but they never surpassed univariate models in number of wins.
Consistent with the full-sample analysis, univariate models are recommended, and among them, the TT task axis is preferable.

\section{Discussion}

\subsection{Findings and Answers to the Research Questions on Normalizing Bias}
\subsubsection{RQ1: univariate $\upsigma$ or bivariate $\upsigma$}
Our answer is to use univariate $\upsigma_\mathit{x}$ for the following reasons.
In the mixed condition, $\id_\xTT$ achieved the highest fit of $R^2=.967537$.
This improved $\approx$.095 points over nominal $\id$ with $R^2=.872731$ (Figures~\ref{fig:resRegressNine}--\ref{fig:resFitMixed}).
All four bivariate models had $R^2<.96$, and their AIC and BIC were worse than the best model by more than 4 points.

For $\tp$ stability, $\id_\xTTAe$ minimized both $\tp_\mathrm{diff}$ and $\tp_\mathrm{cv}$ at 4.74241\% and 2.77263\% (Figure~\ref{fig:resTP}).
Compared to $\id_\xTT$ (the best-fit model), the differences were only 0.08370 and 0.04261 points, respectively.

A simulation further confirmed robustness of univariate models to sample size.
For model fit, bivariate models occasionally won at $N=5$, but at $N \ge 20$ the probability fell to 4.6\%, and at $N \ge 80$ it was zero (Figure~\ref{fig:simCounterFit_R2}).
For $\tp$ stability, the probability that a bivariate model won was $p<.05$ at $N \ge 80$ (Figure~\ref{fig:simCounterTPavg_CV}).

\subsubsection{RQ2: nominal $A$ or $A_\mathrm{e}$}

Our answer depends on the intended metric: use $A$ for model fit and $A_\mathrm{e}$ for $\tp$ stability.
For fit, the $R^2$ difference between the best $\id_\xTT$ and its effective-amplitude version $\id_\xTTAe$ was .002758 (Figure~\ref{fig:resFitMixed}a).
For stability, $\id_\xTTAe$ was best, with differences from $\id_\xTT$ of 0.08370 in $\diff$ and 0.04261 in $\cv$ (Figure~\ref{fig:resTP}b--c).
However, these differences are small, and the simulation results also supported this conclusion.
Thus, there is no reason to avoid either.

\subsubsection{RQ3: TT or CT task axis}

Our answer is to use the TT task axis.
For model fit, the $R^2$ difference between the best model $\id_\xTT$ and the second-best $\id_\xCT$ was negligible ($.000004$, see Figure~\ref{fig:resFitMixed}).
In the simulation, the average $R^2$ difference was .000017 points at $N=5$ (Figure~\ref{fig:simFitR2}), and at $N=160$ the counts of being best were 566 and 434, respectively, roughly a split (Figure~\ref{fig:simCounterFit_R2}).
However, TT never lost to CT on any metric, and thus TT is the safer default.

This conclusion also held for the non-best bivariate models.
For example, for $\cv$ at $N=5$ (Figure \ref{fig:simCounterTPavg_CV}), the combined wins of $\id_\xyTT$ and $\id_\xyTTAe$ were $61+149=210$, while those of $\id_\xyCT$ and $\id_\xyCTAe$ were $46+158=204$, providing no evidence of CT superiority.

\subsection{Additional Support for Using Univariate \texorpdfstring{$\upsigma_\mathit{x}$}{sigma x}}

As Figure~\ref{fig:resMergedNineEndpoints} shows, in ISO-style tasks, endpoints tend to spread along the task axis, and the geometric constraints for hits and misses are mainly determined in that direction.
To align the effective-width principle with this constraint, we should prioritize normalization by $\upsigma_\mathit{x}$.
Specifically, the correction $W_\mathrm{e}=4.133\upsigma$ assumes a 1D normal distribution yielding $\er=3.88\%$.
However, bivariate $\upsigma_\mathit{xy}$ is equivalent to the Euclidean norm of $\upsigma_\mathit{x}$ and $\upsigma_\mathit{y}$: $\upsigma_\mathit{xy}=\sqrt{ \frac{1}{n-1} \sum_{j=1}^{n}  (x_j - \overline{x})^2 + \frac{1}{n-1} \sum_{j=1}^{n}(y_j - \overline{y})^2 }=\sqrt{\upsigma_\mathit{x}^2+\upsigma_\mathit{y}^2}$.
We found no theoretical basis that using $W_\mathrm{e}=4.133\upsigma_\mathit{xy}$ would yield $\er=3.88\%$ or any particular value.

In addition to the tendency for endpoint distributions to elongate along the task axis, most variance due to bias is explained by $\upsigma_\mathit{x}$.
For example, comparing (a) and (g) in Figure~\ref{fig:resMergedNineEndpoints}, when shifting from $\acc$ to $\fast$, $\upsigma_\mathit{x}$ increased from 3.66 to 5.63 pixels (a 53.8\% difference), whereas $\upsigma_\mathit{y}$ increased from 3.21 to 4.34 pixels (35.2\%).
Normalization by $\upsigma_\mathit{xy}$ weights $y$-axis spread equally to the $x$-axis spread (by the above-mentioned Euclidean norm), even though $\upsigma_\mathit{y}$ is comparatively less affected by bias.
This potentially weakens the bias-normalizing effect.
These points align with our empirical finding that $\upsigma_\mathit{x}$ is preferable to $\upsigma_\mathit{xy}$.

\subsection{Why Our Findings Differ from Wobbrock et al.'s Work (2011)}
\label{sec:why}
Wobbrock et al. recommended using $\upsigma_\mathit{xy}$ for 2D pointing \cite{Wobbrock11dim}, whereas our data supported $\upsigma_\mathit{x}$-based models.
Clarifying this divergence helps situate both studies as complementary steps toward a fuller account of the effective-width method.

\subsubsection{Revisiting Wobbrock et al.'s Work}
In their study, 21 participants performed 1D and 2D pointing tasks under a single bias (neutral, targeting 4\% errors).
They computed $W_\mathrm{e}$ with $\upsigma_\mathit{x}$ for 1D and with both $\upsigma_\mathit{x}$ and $\upsigma_\mathit{xy}$ for 2D, then calculated Fitts' law fit in $r$ and $\tp$ per participant for each $\upsigma$ method.
They ran paired $t$-tests three times to compare $r$ values across $\upsigma$ methods.
They found significantly higher mean $r = .962$ in 2D with $\upsigma_\mathit{xy}$ than with $\upsigma_\mathit{x}$ ($.948$ for 1D and $.951$ for 2D).

Our concern is that Fitts' law fits are usually computed on participants' mean $\mt$.
We found no previous studies that computed individual $r$ to test significant differences between $\id$ models.
If three $\upsigma$ computation methods are compared at the individual level, an ANOVA with pairwise tests after $p$-value correction is preferable.
Also, for 2D model comparisons of $\upsigma_\mathit{x}$ vs. $\upsigma_\mathit{xy}$, AIC is an appropriate criterion.

They applied the same $t$-test approach to $\tp$.
In 2D, the mean $\upsigma_\mathit{xy}$ yielded $\tp = 4.49$ bps, which was significantly lower than $\upsigma_\mathit{x}$ in 1D ($4.85$ bps) and 2D ($4.91$ bps).
Because the difference between 1D and 2D using $\upsigma_\mathit{x}$ was only 0.06 bps, they argued $\tp$ was comparable when using $\upsigma_\mathit{x}$.
Again, an ANOVA with $p$-corrected pairwise tests would be appropriate for this purpose.

\subsubsection{Scope and Limitations of Wobbrock et al.'s Work}
Using a single bias leaves the inherent bias-normalizing capability of effective-width method underexplored.
Still, their design filled a gap at the time, since most pre-2011 studies rarely contrasted $\upsigma_\mathit{x}$ and $\upsigma_\mathit{xy}$ even in single-bias settings on metrics such as $r$ and $\tp$.
We therefore regard their choice to begin with the traditional single-bias Fitts' law paradigm, comparing model fit and $\tp$ for $\upsigma_\mathit{x}$ versus $\upsigma_\mathit{xy}$, as a reasonable first step.

One of their conclusions that $\upsigma_\mathit{xy}$ creates a better 2D Fitts' law model makes sense as it incorporates endpoint variation in the direction orthogonal to the task axis and thus can yield a larger $W_\mathrm{e}$.
This, in turn, can produce an $\id_\mathrm{e}$ that more faithfully reflects lower $\tp$ in 2D pointing than in 1D.
Despite our concerns on statistical testing, their empirical observation in 2D that $\upsigma_\mathit{xy}$ produced a higher mean $r = .962$ than $\upsigma_\mathit{x}$ ($.951$) lends support to the plausibility of $\upsigma_\mathit{xy}$ in their setup.

Their experiment used a single bias because earlier studies had scarcely compared $\upsigma_\mathit{x}$ and $\upsigma_\mathit{xy}$ directly.
This design appropriately addressed their research questions such as, ``In a common 2D pointing task, which model yields a higher $r$?''
By contrast, as we argued, their design did not probe the effective-width method's core capability of normalizing across biases.
Hence, both their experiment and ours are best viewed as complementary steps toward understanding the effective-width method.
In actuality, although they recommended $\upsigma_\mathit{xy}$ based on the higher $r$, they did not categorically dismiss $\upsigma_\mathit{x}$, and thus their finding and ours are not directly contradicted.

\subsubsection{Notes to Avoid Misinterpretation}
Because our aims differ from those of Wobbrock et al., it is natural that our design and conclusions also differ.
For example, we do not claim that their finding of higher $r$ for $\upsigma_\mathit{xy}$ was obtained just by chance due to a small sample size (21 participants).
Their experiment itself was properly executed, and within a single-bias 2D setting we find their rationale plausible that $\upsigma_\mathit{xy}$ better captures bivariate endpoint accuracy and yields an $\id_\mathrm{e}$ that predicts $\mt$ more accurately.

Of course, their conclusions might change with larger samples.
For example, the small difference between $r=.962$ using $\upsigma_\mathit{xy}$ and $r=.951$ using $\upsigma_\mathit{x}$ could reverse.
More rigorous statistical procedures might also alter the presence and absence of their reported significant differences.
Likewise, our results could vary under different experimental settings; for example, with a small sample in a laboratory study, $\upsigma_\mathit{xy}$ might outperform in $R^2$ for mixed-bias $\mt$ data.
Given differences in experimental control, instruction adherence, and devices between our study and theirs, the present evidence does not allow us to declare either result right or wrong.
Consistent with prior cautionary notes in the literature \cite{Cockburn20replication,Kasper14once,Yamanaka23chiBias}, we hope future researchers will conduct replications.

\subsection{Implications}

\subsubsection{For Performance Comparison Studies}

Based on our results, we offer the following implications for researchers conducting ISO-style tasks to compare devices, interaction techniques, or user attributes:
\begin{enumerate}
    \item Use $\upsigma_\mathit{x}$. That is, compute effective width from endpoint variability along the direction of motion.
    \item Use $A_\mathrm{e}$ if higher $\tp$ stability is desired.
    \item Define the task axis as the direction from the previous target center to the current target center.
    \item Recruit sufficient participants, ideally more than 20.
\end{enumerate}
Summarizing (1)--(3), we recommend $\id_\xTTAe$ for performance-comparison studies.
This recommendation is based on the stability metrics $\diff$ and $\cv$.
Following these guidelines should reduce $\tp$ variability caused by participants' unconscious speed-accuracy biases \cite{Sharif20,Yamanaka22dis}.

Item (1) matches the recommendations of Soukoreff and MacKenzie \cite{Soukoreff04} and ISO 9241-411 \cite{iso2012}, which might suggest our study finds nothing new.
However, Wobbrock et al. \cite{Wobbrock11ART} later recommended bivariate $\upsigma_\mathit{xy}$.
Many studies after Wobbrock et al.'s work that used $\upsigma_\mathit{xy}$ (see Section~\ref{sec:rwCiteWobbrock2011}) might have compared conditions more fairly if they had used $\upsigma_\mathit{x}$.

For (2), we add the caveat ``if higher stability is desired.''  
Using $A_\mathrm{e}$ agrees with Soukoreff and MacKenzie \cite{Soukoreff04} but contradicts ISO 9241-411's recommendation \cite{iso2012}.
In practice, our analyses show that the choice between $A$ and $A_\mathrm{e}$ barely changes $\tp$.
This supports Zhai et al's claim that the difference between $A$ and $A_\mathrm{e}$ is small \cite{Zhai04speed}.
In research citing Wobbrock et al. \cite{Wobbrock11dim}, both choices appear (Section~\ref{sec:rq2}), and the performance evaluation should not be remarkably affected.

For (3), although rarely examined, we experimentally confirmed that TT and CT differ little.
This is a necessary scientific step, and we show that researchers can safely use the previously recommended TT by default \cite{iso2012,Soukoreff04,Wobbrock11ART}.

For (4), our simulation shows that at $N=20$, bivariate models still often won on $\cv$ ($7+5+83+104=199$ times, $p=.199$; Figure~\ref{fig:simCounterTPavg_CV}c).
Thus, at this sample size, it is still likely to conclude by chance that $\upsigma_\mathit{xy}$ best normalizes bias.
As $N$ increased to 40, 80, and 160, the probabilities decreased to $p=.091$, $.022$, and $.000$, respectively.
As with general guidance for user studies, larger $N$ is essential for reliable results in $\tp$-based performance comparisons.

\subsubsection{Other Implications}

Beyond the above, we offer two additional implications:
\begin{enumerate}
  \setcounter{enumi}{4}
  \item To estimate $\mt$ under the balanced speed-accuracy condition, use the $\id_\xTT$ model.
  \item For reproducibility, when reporting $\tp$ also report the model fit, whether $A$ or $A_\mathrm{e}$ was used, and the definition of task-axis for computing $\upsigma$. 
\end{enumerate}
For (5), previous studies noted that $W_\mathrm{e}$ is determined a posteriori and cannot estimate $\mt$ for new target conditions \cite{Zhai04speed,Wright13}.
However, our results show that $\id_\xTT$ achieves the best $\mt$ predictive accuracy for the mixed condition.
Thus, to estimate $\mt$ at untested conditions when operating to maintain $\er=3.88\%$ (i.e., $W=W_\mathrm{e}$), we recommend $\id_\xTT$.

Item (6) is important for reproducibility.
Some studies reported $\tp$ but not regression results for $\mt$ \cite{MacKenzie15touch,Teather14tilt}.
Because $\tp$ is a metric premised on the assumption that user performance follows Fitts' law, a poor model fit would make the resulting $\tp$ questionable \cite{Yamanaka22chiBias,Yamanaka24MobileHCI}.
In our results, $\id_\xTT$ had slightly lower $\tp$ stability than $\id_\xTTAe$ but higher $R^2$, possibly yielding more trustworthy $\tp$.
Currently, when fit and stability identify different best models, there is no guidance for which to choose. We leave this for future work.
In practice, since the differences are small and $A_\mathrm{e}$ adjusts even if the actual movement distance changes, $\id_\xTTAe$ is a safe choice.

\subsection{Limitations and Future Work}

Our study has limitations, such as using only mice.
Other devices or VR systems might yield larger endpoint variability orthogonal to the task axis and thus bivariate models might perform differently.
Crowdsourced studies also face uncontrolled factors affecting pointing performance, such as display and mouse disparities, mouse-to-cursor latency \cite{Casiez11,Casiez15}, and control-display gain \cite{Pang19gain,Lee20autogain}.

A further limitation lies in the relatively low task difficulty of $\id = $ 2.07--4.70 bits compared with prior work (e.g., 1.58--6.02 bits in Wobbrock et al.'s study \cite{Wobbrock11dim}).
However, many HCI studies have also used similarly low $\id$ ranges or employed crowdsourcing.
Based on this precedent, our results still provide valuable contributions to the HCI field.
For example, while Wobbrock et al.'s study did not fully evaluate the inherent capability of the effective-width method, we introduced three bias conditions that produced large effects on $\mt$ and $\er$ ($\upeta_\mathrm{p}^2=.5815$ and $.7011$, respectively, see Tables~\ref{tab:resAnovaMT} and \ref{tab:resAnovaER}) and thoroughly analyzed which $\upsigma$ computation best normalized those effects.
While our study shares limitations common to many others, it nevertheless provides novel and useful insights that prior work has not previously demonstrated.
Future replications employing a broader $\id$ range or laboratory settings would add further contributions, but our study still maintains the internal validity.

Previous studies used task settings whereby participants were not required to re-select the current target after an error \cite{iso2012,Wobbrock11ART}.
In such cases, they would be more inaccurate and thus the first click in each trial may miss the target more than our current study.
This would result in clearer observations of the effects of $A_\mathrm{e}$ and the task-axis definition.
If future researchers aim to focus on comparisons under these conditions rather than the effective width, separate experiments should be specifically designed for that purpose.

Enforcing stricter instruction compliance is another direction.
For example, using nominal $\id$ under the mixed condition, Zhai et al. showed $R^2=.696$ \cite{Zhai04speed} and Yamanaka showed $R^2=.6094$ \cite{Yamanaka24ijhcimerit}, whereas we showed higher $R^2$ of .872731 (Figure~\ref{fig:resRegressNine}a).
Our participants may have differentiated the three biases less than in those studies.
One possible reason behind the relatively smaller bias effect was that requiring the participants to re-aim for the target when errors occurred shifted their unconscious bias towards accuracy to save the time and effort.
To strengthen instruction compliance, one could set explicit goals like MacKenzie and Isokoski who ensured that $\mt$ differences across biases were greater than 10\% \cite[p.~1634]{MacKenzie08} or use metronomes to impose temporal constraints \cite{Wobbrock08error}.
Even so, our data showed $\bias$ had large effect sizes on $\mt$ and $\er$, which indicated sufficient compliance.
Given these substantial differences, our dataset is appropriate for comparing different $\upsigma$-calculation methods to determine which can best normalize the bias effects.
Furthermore, our conclusions were also supported by the 1{,}000-iteration simulation, which included many cases with larger separations in $\mt$ among the three biases.

We considered filtering to include only participants with perfectly ordered $\mt_\acc > \mt_\neu > \mt_\fast$ and $\er_\acc < \er_\neu < \er_\fast$ per bias.
However, because errors occur stochastically, the ideal order may not hold with only a few trials per condition (25 in our study).
Randomized bias-order per participant may also introduce small learning effects that might affect this ordering for some participants.
Our focus was not on individual-level analysis but on identifying models that best normalize bias for typical user studies, such as comparing mean $\tp$s across devices or other conditions.
Future laboratory experiments can test whether our findings replicate under stricter control to strengthen our conclusions.

\section{Conclusion}

We aimed to clarify confusion around the calculation of effective target width for 2D pointing and to provide procedures and criteria for future standardization of measurement and reporting.
To that end, we compared $\ide$ models that normalize speed-accuracy bias in an ISO-style pointing task from theoretical, experimental, and simulation perspectives.
In the mixed condition, where data from three speed-accuracy bias conditions were combined, the univariate model $\id_\xTT$ achieved the highest $R^2$; the bivariate models were inferior in AIC and BIC.
$\tp$ stability was best with $\id_\xTTAe$.
Comparisons of $A$ vs. $A_\mathrm{e}$ as well as TT vs. CT task axes had small effects on both model fit and $\tp$ stability, so $A_\mathrm{e}$ and TT can serve as the practical default.
The simulation showed that these conclusions are robust for small to medium sample sizes, and that the utility of bivariate models drops rapidly for $N>20$.
We therefore recommend $\id_\xTTAe$ to enable fair performance comparisons in future studies.


\bibliographystyle{ACM-Reference-Format}
\bibliography{sample-base}

\appendix
\section{Research Ethics Approval}
\label{sec:appendixEthics}
This study was approved by our affiliation's IRB-equivalent research ethics committee (including the legal department) and the crowdsourcing platform.
We obtained informed consent from all workers, who agreed that we record and publish data to the extent necessary for research.
The committee's decision is that sufficiently summarized data (e.g., mean $\mt$) can be published, but analyses outside the research purpose (e.g., relating individual age or gender to performance) are not permitted.
As in previous Fitts' law studies, we mainly analyze model fit based on mean $\mt$ and mean $\tp$, so this decision does not affect our conclusions.

\end{document}